\documentclass[a4paper,fuseAMS, leqn, usenatbib]{mnras}
\let\pwiflocal=\iffalse \let\pwifjournal=\iffalse
\usepackage[utf8]{inputenc}
\usepackage{amsmath,amssymb,gensymb,times,graphicx,morefloats,color}
\usepackage{wasysym}
\usepackage{textcomp}
\usepackage{afterpage, bm}
\usepackage{natbib,hyperref}
\usepackage{wrapfig}
\usepackage{float}

\usepackage{ulem}
\usepackage{xcolor}
\usepackage[outdir=./]{epstopdf}
\usepackage{mathrsfs}
\usepackage{mathtools}
\usepackage{url}

\usepackage{subfigure}
\usepackage{float}
\usepackage{longtable}
\usepackage[T1]{fontenc}
\usepackage{ae,aecompl}

\usepackage{listings}
\definecolor{dkgreen}{rgb}{0,0.6,0}
\definecolor{gray}{rgb}{0.5,0.5,0.5}
\definecolor{mauve}{rgb}{0.58,0,0.82}
\definecolor{golden}{rgb}{0.86,0.65,0.01}
\newcommand{\teff}{T$_{\mathrm{eff}}$}
\newcommand{\logg}{log $g$}
\newcommand{\feh}{[Fe/H]}

\newcommand{\vsini}{V~$\sin{i}$}

\title[Late-type HVS candidates]{A Detailed Chemical Study of the Extreme Velocity Stars in the Galaxy}
\author[Nelson et al. 2024]{Tyler~Nelson$^{1}$\thanks{E-mail:tyler.w.nelson@maine.edu}, Keith~Hawkins$^{2}$, Henrique~Reggiani$^{3}$, Diego~Garza$^{4}$, Rosemary~F.G.~Wyse$^{5}$, Turner~Woody$^{6}$\\
$^{1}$Department of Physics, University of Southern Maine, 96 Falmouth Street, Portland, ME 04103, USA\\
$^{2}$Department of Astronomy, The University of Texas at Austin, 2515 Speedway Boulevard, Austin, TX 78712, USA\\
$^{3}$The Observatories of the Carnegie Institution for Science, 813 Santa Barbara St, Pasadena, CA 91101, USA\\
$^{4}$Department of Astronomy \& Astrophysics, The University of California at Santa Cruz, 1156 High Street, Santa Cruz, CA 95064, USA\\
$^{5}$Physics and Astronomy Department, Johns Hopkins University, 3400 North Charles Street, Baltimore, MD 21218, USA\\
$^{6}$Harvard-Smithsonian Center for Astrophysics, Cambridge, MA, 02138, USA
}

\date{Accepted 2024 June 26. Received 2024 May 21; in original form 2023 September 29}

\pubyear{2024}

\begin{document}
\label{firstpage}
\pagerange{\pageref{firstpage}--\pageref{lastpage}}
\maketitle

\begin{abstract}

Two decades on, the study of hypervelocity stars is still in its infancy. These stars can provide novel constraints on the total mass of the Galaxy and its Dark Matter distribution. However how these stars are accelerated to such high velocities is unclear. Various proposed production mechanisms for these stars can be distinguished using chemo-dynamic tagging. The advent of Gaia and other large surveys have provided hundreds of candidate hyper velocity objects to target for ground based high resolution follow-up observations. We conduct high resolution spectroscopic follow-up observations of 16 candidate late-type hyper velocity stars using the Apache Point Observatory and the McDonald Observatory. We derive atmospheric parameters and chemical abundances for these stars. We measure up to 22 elements, including the following nucleosynthetic families: $\alpha$ (Mg, Si, Ca, Ti), light/Odd-Z (Na, Al, V, Cu, Sc), Fe-peak (Fe, Cr, Mn, Co, Ni, Zn), and Neutron Capture (Sr, Y, Zr, Ba, La, Nd, Eu). Our kinematic analysis shows one candidate is unbound, two are marginally bound, and the remainder are bound to the Galaxy. Finally, for the three unbound or marginally bound stars, we perform orbit integration to locate possible globular cluster or dwarf galaxy progenitors. We do not find any likely candidate systems for these stars and conclude that the unbound stars are likely from the the stellar halo, in agreement with the chemical results. The remaining bound stars are all chemically consistent with the stellar halo as well.

\end{abstract}


\section{Introduction}
\label{introduction}

High-velocity (HiVel) stars are unique dynamical probes for understanding the Galaxy \citep[e.g.,][]{hawkins_2015_hivel}. Gravitationally bound HiVel stars can be used to constrain the total mass and local escape velocity of the Galaxy \citep[e.g.,][]{Piffl_2014, Williams_2017}. Unbound HiVel stars can be used to study the the Galaxy's dark matter halo \citep[e.g.,][]{Gnedin_2005, Gallo_2022}. The acceleration mechanisms for these unbound HiVel stars remains unclear \citep[see e.g.,][and references therein]{Tutukov_2009, Brown_2015}. 

Unbound HiVel stars, which we will refer to as hypervelocity stars\footnote{This definition is agnostic of production mechanisms for the unbound stars. Some authors use hypervelocity star to label objects from the Hills mechanism, and runaway/hyper-runaway stars for other fast moving stars not produced in this manner. Under this alternative definition, there is then a discussion of bound and unbound hypervelocity stars. Other authors have adopted high velocity and extreme velocity to be agnostic to the origin/production mechanisms.} (HVSs), were first proposed by \cite{Hills_1988} with a more narrow usage. \cite{Hills_1988} defined HVSs as unbound stars moving on radial orbits from the Galactic Center (GC), potentially having galactocentric rest frame velocities $v_{\mathrm{GRF}}>1000 \ \mathrm{km \ s^{-1}}$. They argued these stars were the product of a 3-body encounter consisting of a stellar binary and a super massive black hole (SMBH). This production pathway is the so-called ``Hills' mechanism''. However, there are myriad potential origins for HVSs stars because of the broad definition we adopt, including accreted systems \citep{Henrique_2022}, the stellar disc, and the stellar halo, among others \citep[see e.g.,][and references therein]{quispe_2022}. HVSs can constrain the total mass of the Galaxy \citep{rossi_2017}, and the environment at the GC \citep[][]{kenyon08, Brown_2015, rossi_2017, marchetti_2022}. Furthermore, some models for HVS production from the Large Magellanic Cloud (LMC) provide indirect evidence for the existence of either a massive black hole \citep{lmc_hvs_smbh, lmc_hvs_prediction} or an intermediate mass black hole \citep{imbh_lmc_hvs} at the center of the LMC.

\cite{brown_2005} provided the first observational evidence for the Hills' mechanism. They observed a B-type star (labeled HVS1) with $v_{\mathrm{GRF}} \sim 673 \ \mathrm{km \ s^{-1}}$ and a galactocentric distance of $107 \ \mathrm{kpc}$ \citep{Brown_2014}. This is often claimed to be the first HVS observed; however, this depends on the definition of HVS being used. This serendipitous discovery and the numerous large scale surveys of the Milky Way's stellar populations, e.g. The Radial Velocity Experiment \citep[RAVE,][]{rave_paper}, The Apache Point Observatory Galactic Evolution Experiment \citep[APOGEE,][]{APOGEE_2017}, The Sloan Digital Sky Survey \citep[SDSS,][]{sloan_paper}, Gaia \citep{Gaia_DR2}, have caused dramatic growth in the study of HiVel stars. Much of this investigation is focused on HVSs \citep[reviewed in][]{Brown_2015}, with less attention paid to the bound stars. However, recent works have shed light on these bounded stars as well \citep[e.g.,][]{hawkins_2015_hivel, Hawkins_2018, Henrique_2022, quispe_2022}. 

The Hills' mechanism alone cannot explain all the unbound stars observed in the Galaxy. \cite{heber08} showed that the B-type star HD 271791 could not have originated from the GC because its flight time would be at least twice the lifespan of the star. In addition, the apparent clumping of early type HVSs around the constellation Leo \citep[][]{Brown_2014, Brown_2015} does not agree with the expectation that HVSs from the Hills' mechanism should be isotropically distributed around the GC. Hence competing ideas on production emerged. In addition, there were variations on the Hills' mechanism which could produce HVS \citep[e.g., a star interacting with a massive black hole binary,][]{Yu_2003}. Runaway stars are one such idea, where the observed star was jettisoned from its birth star cluster and accelerated to high velocities. This could be accomplished through dynamical evolution in clusters \citep{Poveda67}, binary interactions \citep{Leonard88} or binary supernova explosion \citep{Blaauw61}. Another possibility is the so-called double degenerate double detonation scenario, where two white dwarfs orbit each other, the primary star undergoes a helium shell detonation and a subsequent carbon core detonation in a type 1a supernova. Afterwards, the secondary white dwarf is accelerated to high velocity from the resulting explosion. This mechanism has been suggested for three HVS white dwarfs observed by \cite{Shen18} and six HVS white dwarfs observed by \cite{ElBadry2023}. Other origins include tidal stripping of globular clusters \citep[][]{capuzzo_2015}, satellite dwarf galaxies \citep[e.g.,][]{Pereira12, lmc_hvs_prediction}, or merging galaxies \citep[e.g.,][]{Abadi09, Pereira12, helmi18}. These HiVel stars could also originate in the stellar halo and be subsequently dynamically heated through a merger event. We refer the reader to \cite{Tutukov_2009} and \cite{Brown_2015} and references therein for a more comprehensive list of possible acceleration mechanisms. Even more production pathways exist for bound HiVel stars because the energies required are less extreme compared to the unbound stars. These production pathways for HVSs are often difficult to distinguish from one another entirely; however, some progress can be made studying the possible origins of observed HVSs and their spatial distributions across the Galaxy \citep{Brown_2015, Hawkins_2018}.

The small sample size of confirmed HVSs is a fundamental barrier to both disentangling the plethora of formation pathways proposed for these stars and their application to study the Galaxy. The small sample is both a property of their intrinsic rarity and our ability to detect these stars. The review by \cite{Brown_2015} estimates the sample size of confirmed HVS is $\sim20$ based on prior literature. Distinguishing production mechanisms on the basis of ejection velocity would require 50-100 \citep{sesana_07, Perets_09} HVSs. Applications of HVSs also can require much larger samples \citep[e.g.,][requires up to 800 HVSs to constrain the DM halo shape]{Gallo_2022}. Many studies have produced candidate HVSs following the discovery in \cite{brown_2005}. \cite{Boubert_2018} compiled a catalog of HVS candidates in the literature, finding over 500. \cite{Boubert_2018} re-examined this catalog of candidates using Gaia DR2 measurements whenever possible, because of the uniform treatment of data and the improvements in astrometric precision, finding $N\sim40$ had a probability of being bound to the Galaxy below 50\%. This sample only had 1 late type star\footnote{i.e., spectral type FGKM} present. A slew of new HVS candidates have been discovered following Gaia DR2 and DR3 \citep[e.g.,][]{Bromley_2018, Hattori_2018a, Marchetti_2019b, Raddi_2021, Li_2021, Marchetti_2021, Igoshev_2023}, the majority of which are oriented towards either late type stars or white dwarfs, which had not been readily sampled before Gaia DR2 \citep[see e.g.,][]{Boubert_2018}. These developments in turn have spurred interest in characterizing these candidate HVSs and other HiVel stars \citep[e.g.,][]{Hawkins_2018, Henrique_2022, quispe_2022}. These studies use chemo-dynamic approaches to constrain the origins of these candidate HVSs. Regardless of whether the objects are truly bound or not, constraining the origin of the sample of HVS candidates is interesting because of the diverse range of phenomena that can produce these HiVel stars. \cite{Hawkins_2018} finds their sample is comprised of halo stars, while \cite{Henrique_2022} and \cite{quispe_2022} find large fractions ($\sim50\%$, and $\sim86\%$, respectively) of their samples are consistent with an accreted origin.

This study aims to expand the number of well characterized extreme velocity stars using candidates from the literature, in a similar vein as \cite{Hawkins_2018} and \cite{Henrique_2022}. We set out to take ground based observations of 16 candidate hyper velocity stars to more precisely constrain their radial velocities. We then chemically characterize them so that we may place constraints on their likely origin.  This chemical characterization has seen success in \cite{Hawkins_2018} and \cite{Henrique_2022}. With our sample of 16 stars, we substantially enlarge the pool of extreme velocity stars with chemical abundances. In Section \ref{target_selection} we summarize our target selection. Section \ref{observations} details the data acquisition and reduction. Our methods for measuring the atmospheric parameters and chemical abundances are provided in Sections \ref{atmospheric_section} and \ref{section:abund_methods} respectively. A description of the kinematic analysis is given in Section \ref{kinematic_section}. The results are presented in Section \ref{sec:results} and discussed in Section \ref{discussion}. Finally a summary is given in Section \ref{summary}.

\section{Data Properties}
\subsection{Target Selection}
\label{target_selection}

The goal of this work is to constrain the origins and production mechanisms for these HVS candidates. In order to achieve this goal, we start by selecting HVSs to follow up from various existing literature sources \citep{Bromley_2018, Hattori_2018a, Marchetti_2019b}, and from Astronomical Data Query Language (ADQL) queries by the authors using Gaia DR2/DR3 data shown in Appendix \ref{appendix:adql_query}. Each method uses different selection criteria therefore we will summarize each. For brevity, we omit the various quality cuts imposed by each study and encourage the interested reader to see the original work for more details.

\cite{Bromley_2018} and \cite{Marchetti_2019b} used three dimensional (3D) velocities and orbit integration with a Milky Way (MW) gravitational potential. The two studies differ in selection criteria and the masses used for the MW potential (we refer the reader to Section 2.5 of \cite{Bromley_2018} for more details on the differences between the works) and consequently may find different HVS candidates. \cite{kenyon_2018_comp_rv_vt} have found radial and tangential velocities can be used in lieu of full 3D velocities as a reliable method for finding HVSs depending on the star's distance from the Sun. Tangential velocities are more useful for nearby stars (i.e., $\lesssim 10$ kpc from the Sun) because of the lower uncertainties in parallax, while radial velocities (RVs) are useful at further distances. \cite{Hattori_2018a} finds 30 candidate HVSs within 10 kpc from the Sun using only the tangential velocities. \cite{Hattori_2018a} note that their approach is complementary with \cite{Marchetti_2019b}, as they use different quality cuts on the astrometric data, allowing them to potentially sample a different group of stars. Finally, we have found candidates based on galactocentric radial velocities from Gaia DR3 data. The coordinate system transformations were done using Equation 1 from \cite{hawkins_2015_hivel}.

\subsection{Follow-up Observations}
\label{observations}
The final target selection consisted of 16 late-type HVS candidates predominately in the northern hemisphere. This sample complements the data from \cite{Henrique_2022}, who used a similar sample size of candidate HVSs in the southern hemisphere to study said stars' chemistry. In addition to the program stars, we observe stars from the Gaia Benchmark catalog \citep{benchmark_temperature_grav, benchmark_metal, benchmark_library} and \cite{Bensby_2014} catalog. These standard stars assist in refining the data reduction, verifying the data analysis, and calibrating derived abundances. Lastly, we reanalyze some data from \cite{Hawkins_2018} and \cite{Henrique_2022} to assess the impact of methodological differences between the studies. 

High-resolution spectra were collected using two instruments: the ARC Echelle Spectrograph (ARCES) on the 3.5m Apache Point Observatory Telescope \citep{ARCES_paper}, and the Tull Echelle Spectrograph \citep[TS,][]{TS_paper} on the 2.7m Harlan J. Smith Telescope at the McDonald Observatory. ARCES observations completely sample $3800-9200$ \AA\ with a resolving power $R = \lambda/\Delta \lambda \sim 31500$. TS observations used slit 4 with a resolving power of $\sim60000$ and a wavelength coverage of $\sim 3500 - 10000$ \AA\, with interorder gaps towards the redder wavelengths. For both instruments, standard calibration exposures were also obtained (i.e., biases, flats, and ThAr lamp). Raw data was reduced in the usual fashion (i.e., bias removal, flat-fielding, cosmic ray removal, scattered light subtraction, optimal extraction, and wavelength calibration) using pyRAF/IRAF\footnote{IRAF is distributed by the National Optical Astronomy Observatory, which is operated by the Association of Universities for Research in Astronomy (AURA) under a cooperative agreement with the National Science Foundation.}. 

To normalize the spectra, we fit a pseudo continuum using cubic splines and iterative sigma clipping. Orders are then combined using a flux weighted average. We discard 50 pixels on either end of each order because of the poor signal due to the blaze function. We compared the normalization of the Gaia Benchmark stars we observed to a reference normalization \citep{benchmark_library} to fine tune the sigma clipping parameters. The signal-to-noise ratio (SNR) was estimated at the end of the normalization and order stitching by calculating the standard deviation in the normalized flux with values between 1 and 1.2 over a 60 \AA \ window\footnote{The width of the wavelength window is roughly half the wavelength range of an order and therefore gives a middle ground as to the quality of data on average.} in the middle of the chip at 5200 \AA. Assuming Gaussian noise, we can transform this into a robust estimate for the true noise using a half-normal distribution\footnote{The choice of a half-normal distribution was motivated by the relative ease to measure noise above the continuum, compared to below the continuum where absorption features must be contended with.}. The upper limit on the flux mitigates the impact of hot pixels and cosmic rays\footnote{The choice of an upper limit to remove spurious large flux values could inflate the SNR. In practice, the influence of this choice only changes the SNR by a few for most stars.}, while the lower limit avoids confusing absorption features with noise. Our median SNR over the range $5170-5230$ \AA \ was 28 per pixel for ARCES, 32 per pixel for TS. \texttt{iSpec} \citep{ispec} was used to perform the final bad pixel removal\footnote{To remove the influence of hot pixels we masked that data out and inflated the errors on the points around them by a factor of 10. For dead pixels, we masked but did not inflate the errors of neighboring data.}, radial velocity corrections and the barycentric correction. The radial velocity was determined using a cross correlation with an atomic line list. These results agreed very well with the Gaia DR3 radial velocities. The barycentric corrections were found using the built in iSpec calculator. A summary of the observational parameters can be found in Table~\ref{tab:obs}. Data from \cite{Hawkins_2018} and \cite{Henrique_2022} was processed in an identical manner.

Two targets, Star 1 (Gaia DR3 source~id 1400950785006036224) and Star 6 (Gaia DR3 source~id 1042515801147259008) were observed with both ARCES and TS providing a check on the validity of our reduction method. 

\begin{figure}
    \centering
    \includegraphics[width=1\columnwidth]{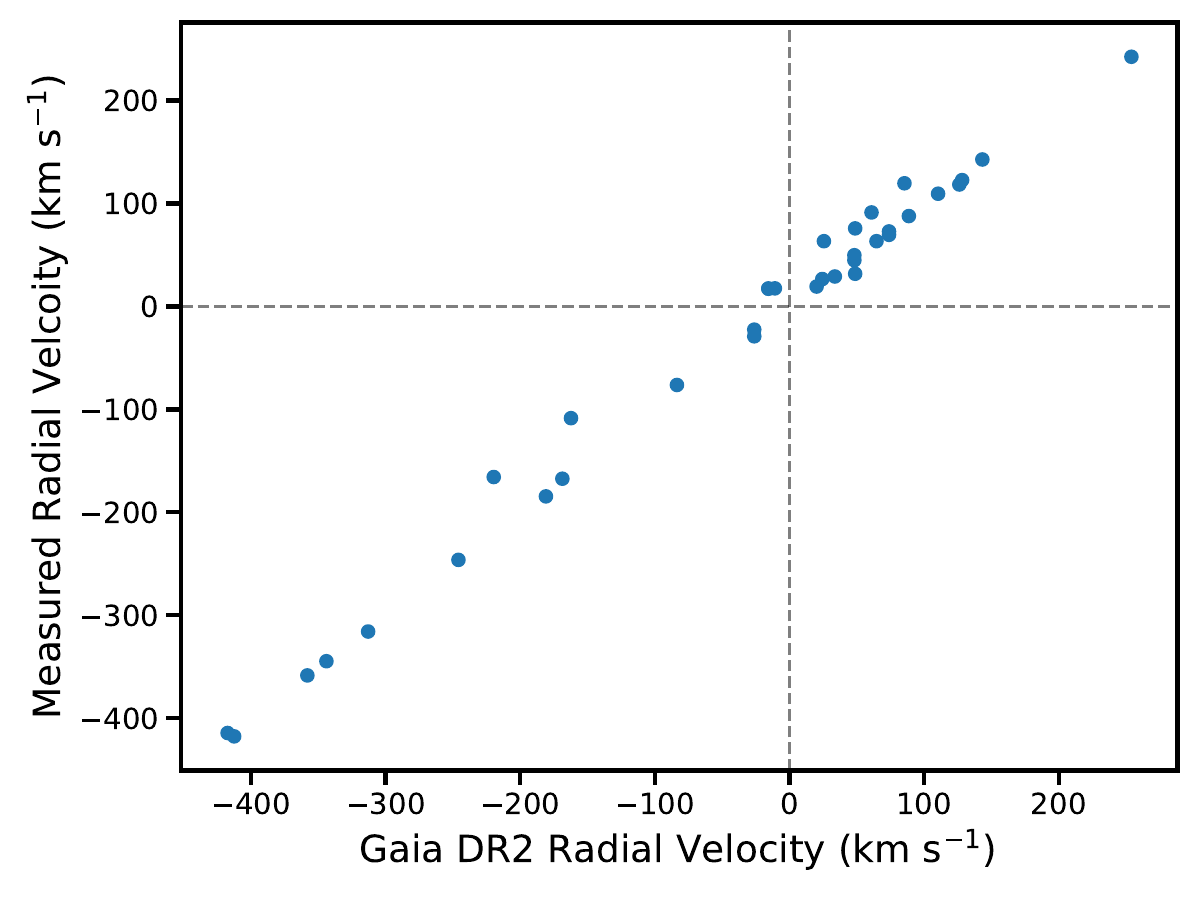}
    \caption{Displayed is a comparison of the radial velocities measured in this study versus those from Gaia DR2. Generally we find very good agreement between the two studies. Error bars are included and are smaller than the typical point size. Dashed lines indicate a radial velocity of 0 (km s$^{-1}$).} 
    \label{fig:rv_fig}
\end{figure}

The initial sample had a contaminant (Gaia DR3 source~id 4150939038071816320). We believe this was from problems with the observed spectrum from the first version of Gaia DR2 leading to an erroneously large RV measurement, with $|\mathrm{V_{r}}| > 500 \ \mathrm{km \ s^{-1}}$. Our RV measurements indicate this is not an extreme velocity star, $\mathrm{V_{r}} \sim -21.6\pm1 \ \mathrm{km \ s^{-1}}$ which agrees with the Gaia DR3 estimate of $\mathrm{V_{r}} \sim -22.3\pm 3.9 \ \mathrm{km \ s^{-1}}$, with a total velocity similar to the Sun. We conclude it is not an extreme velocity star and is omitted from our data tables. It was processed in the same manner as the science sample and provides another check on our methodology.  We compared our radial velocity measurements with the estimates from Gaia in Figure \ref{fig:rv_fig}. We find good agreement between the Gaia DR2 estimates and our measurements, with the Gaia measurements being on average 7.5 km s$^{-1}$ larger than the values we measured from our follow up observations.

\begin{table*}
\caption{The observational parameters of the stars used in this study. A complete machine-readable version is available online. The astrometry is from Gaia DR3. We elect to use the distances from \protect\cite{Bailer-Jones_2021} in lieu of the values from parallax inversion from Gaia DR3 because more than a third of our science stars have relative parallax errors greater than 10\%. Radial velocities are derived from the ground based follow-up observations along with the signal-to-noise ratio (SNR). The literature source of the candidate hyper velocity star is provided in the reference column. Stars used for calibrations or previous detailed in other studies \protect\citep[e.g.,][]{Hawkins_2018,Henrique_2022} are not included in our data tables. We use the abbreviation McD to indicate the observations were taken at the McDonald Observatory, and APO for the Apache Point Observatory. The ADQL entry in the source column indicates the targets were acquired from the ADQL query listed in Appendix \ref{appendix:adql_query}.}
\label{tab:obs}
\centering
\begin{tabular}{ccccccccccc}
\hline
\hline
Gaia DR3 source id & Alias & Observatory & RA & DEC & Distance & G & RV & $\sigma_{\mathrm{RV}}$ & SNR & Reference \\
 &  &  & $\mathrm{{}^{\circ}}$ & $\mathrm{{}^{\circ}}$ & $\mathrm{pc}$ & $\mathrm{mag}$ & km s$^{-1}$ & km s$^{-1}$ &  &  \\
 \hline
1400950785006036224 & star 1 & McD & 233.9279 & 46.5688 & 5582 & 13.07 & 49.7 & 0.2 & 28 & \cite{Hattori_2018a} \\
4531575708618805376 & star 2 & APO & 281.8599 & 22.1394 & 6322 & 13.04 & -420.1 & 0.1 & 28 & \cite{Marchetti_2019b} \\
2629296824480015744 & star 3 & APO & 335.8334 & -2.5197 & 638 & 11.36 & -165.6 & 0.2 & 22 & \cite{Hattori_2018a} \\
3784964943489710592 & star 4 & APO & 169.3563 & -5.8154 & 2555 & 12.24 & 118.4 & 0.2 & 66 & \cite{Marchetti_2019b} \\
1396963577886583296 & star 5 & APO & 237.7316 & 44.4357 & 19923 & 13.23 & -417.5 & 0.1 & 18 & \cite{Marchetti_2019b} \\
1042515801147259008 & star 6 & McD & 129.7990 & 62.5013 & 2110 & 12.71 & 72.9 & 0.5 & 36 & \cite{Hattori_2018a} \\
1383279090527227264 & star 7 & APO & 240.3373 & 41.1668 & 6311 & 13.00 & -184.4 & 0.2 & 46 & \cite{Bromley_2018} \\
1478837543019912064 & star 8 & McD & 212.4777 & 33.7129 & 5805 & 13.08 & -246.1 & 0.6 & 27 & \cite{Bromley_2018} \\
1552278116525348096 & star 9 & McD & 204.6690 & 48.1565 & 1603 & 12.66 & -76.3 & 0.3 & 43 & \cite{Hattori_2018a} \\
3705761936916676864 & star 10 & McD & 192.7642 & 4.9411 & 2836 & 13.18 & 87.8 & 0.2 & 13 & \cite{Hattori_2018a} \\
4395399303719163904 & star 11 & McD & 258.7501 & 8.7314 & 6591 & 13.17 & 26.6 & 0.2 & 26 & \cite{Marchetti_2019b} \\
1598160152636141568 & star 12 & McD & 234.3616 & 55.1622 & 3063 & 10.78 & -167.3 & 0.2 & 47 & \cite{Hattori_2018a} \\
4535258625890434944 & star 13 & McD & 278.4475 & 23.1167 & 3804 & 13.13 & -204.4 & 0.6 & 27 & \cite{Bromley_2018} \\
2159020415489897088 & star 14 & McD & 273.3214 & 61.3187 & 4893 & 12.50 & -108.5 & 0.6 & 32 & \cite{Bromley_2018} \\
3713862039077776256 & star 15 & McD & 206.5166 & 4.1533 & 4798 & 11.17 & 489.8 & 0.1 & 42 & ADQL Query \\
4531308286776328832 & star 16 & McD & 282.5286 & 21.6281 & 2582 & 11.83 & -619.2 & 0.1 & 54 & ADQL Query \\
\hline
\end{tabular}
\end{table*}

\subsection{External Data}
\label{external_data}
We use external data for our targets to aid in the isochrone analysis in Section \ref{sec:isochrones} and the kinematic analysis in Section \ref{kinematic_section}. We use astrometric data from Gaia DR3 \cite{Gaia_EDR3_summary}. Rather than using parallax to estimate distance, we use the distance estimates from \cite{Bailer-Jones_2021} because more than a third of our stars have relative parallax errors greater than 10\%. 

We use the following photometric data (when available): 
\begin{enumerate}
    \item Gaia DR2 $G$ band magnitude \citep{Gaia_mission, Gaia_DR2_summary, Gaia_DR2_photometry}
    \item 2MASS $J, H, K_s$ bands and associated uncertainties \citep{2MASS_survey}
    \item AllWISE W1, and W2 bands and associated uncertainties \citep{wise_1}
    \item SkyMapper u, v, g, r, i, z bands and associated uncertainties \citep{SkyMapper}
    \item PANSTARRs g, y bands and associated uncertainties \citep{panstarrs_survey, panstarrs_photometry}
    \item SDSS u, z bands and associated uncertainties \citep{sdss_4_summary}
\end{enumerate}

These bands must pass quality cuts\footnote{SkyMapper: \href{http://skymapper.anu.edu.au/data-release/}{link}\\
SDSS: \href{https://www.sdss.org/dr13/tutorials/flags/}{link 1},\href{http://skyserver.sdss.org/dr13/en/help/docs/realquery.aspx\#cleanStars}{link 2}\\
Pan-STARRS: \href{https://outerspace.stsci.edu/display/PANSTARRS/Pan-STARRS1+data+archive+home+page}{link 1},\href{https://outerspace.stsci.edu/display/PANSTARRS/PS1+Sample+queries\#PS1Samplequeries-Gethigh-fidelitystarsingivenarea}{link 2}, \href{https://outerspace.stsci.edu/display/PANSTARRS/PS1+Object+Flags}{link 3}\\
2MASS: \href{https://irsa.ipac.caltech.edu/data/2MASS/docs/releases/allsky/doc/sec1\_6b.html}{link 1}, \href{https://irsa.ipac.caltech.edu/data/2MASS/docs/releases/allsky/doc/sec2\_2a.html}{link 2}}. These cuts are identical to the recommendations put forward by each survey, with the exception of SDSS where we allowed a bad pixel within 3 pixels of the centroid, otherwise none of our stars would have usable SDSS photometry. Since the SDSS u-band is the bluest band we use, retaining it is important for constraining the metallicity and extinction. All photometric data used in our subsequent fitting is provided in Table \ref{tab:photometry}. We also use extinction estimates from \texttt{BAYESTARS} \citep{Green_2019}.

\begin{table*}
\centering
\caption{A portion of the photometric Data used for isochrone fitting. For compactness, we only display a subset of the columns. We assume an error of 0.005 mags for the Gaia G band magnitude. When values are not present or do not pass our quality cuts, a nan value is provided. The Gaia photometry corresponds to Gaia DR2 with the values from Gaia DR3 producing no changes. The $J$, $\sigma_{J}$, $H$, $\sigma_{H}$, $K_s$, $\sigma_{K_{s}}$ correspond to the 2MASS survey. $W1$ and $W2$ are photometry from AllWISE. We use SkyMapper DR2 data, PANSTARRs DR1 data, and SDSS IV data when available. A full machine readable version of the table is available online.  \label{tab:photometry}}
\resizebox{\textwidth}{!}{%
\begin{tabular}{ccccccccccccc}
\hline
\hline
Alias   & Gaia DR3  & Gaia G         & $J$            & $\sigma_{J}$   & $H$            & $\sigma_{H}$   & $K_s$          & $\sigma_{K_{s}}$ & $W_1$          & $\sigma_{W_{1}}$ & $W_2$          & $\sigma_{W_{2}}$ \\
        &                     & $\mathrm{mag}$ & $\mathrm{mag}$ & $\mathrm{mag}$ & $\mathrm{mag}$ & $\mathrm{mag}$ & $\mathrm{mag}$ & $\mathrm{mag}$   & $\mathrm{mag}$ & $\mathrm{mag}$   & $\mathrm{mag}$ & $\mathrm{mag}$   \\
        \hline
star 1  & 1400950785006036224 & 13.073         & 11.53          & 0.02           & 10.91          & 0.02           & 10.88          & 0.01             & 10.79          & 0.02             & 10.84          & 0.02             \\
star 2  & 4531575708618805376 & 13.043         & 11.14          & 0.02           & 10.50          & 0.02           & 10.38          & 0.02             & 10.26          & 0.02             & 10.33          & 0.02             \\
star 3  & 2629296824480015744 & 11.363         & 9.93           & 0.02           & 9.51           & 0.02           & 9.38           & 0.02             & 9.31           & 0.02             & 9.35           & 0.02             \\
star 4  & 3784964943489710592 & 12.238         & 10.71          & 0.03           & 10.18          & 0.02           & 10.10          & 0.02             & 10.02          & 0.02             & 10.08          & 0.02             \\
star 5  & 1396963577886583296 & 13.229         & 11.13          & 0.02           & 10.36          & 0.02           & 10.19          & 0.02             & 10.11          & 0.02             & 10.19          & 0.02             \\
star 6  & 1042515801147259008 & 12.706         & 11.13          & 0.02           & 10.68          & 0.03           & 10.59          & 0.02             & 10.51          & 0.02             & 10.52          & 0.02             \\
star 7  & 1383279090527227264 & 12.998         & 11.52          & 0.02           & 10.99          & 0.02           & 10.90          & 0.02             & 10.82          & 0.02             & 10.86          & 0.02             \\
star 8  & 1478837543019912064 & 13.083         & 11.75          & 0.02           & 11.31          & 0.02           & 11.22          & 0.02             & 11.17          & 0.02             & 11.18          & 0.02             \\
star 9  & 1552278116525348096 & 12.664         & 11.52          & 0.02           & 11.14          & 0.03           & 11.11          & 0.02             & 11.04          & 0.02             & 11.07          & 0.02             \\
star 10 & 3705761936916676864 & 13.184         & 11.77          & 0.02           & 11.27          & 0.03           & 11.21          & 0.02             & 11.13          & 0.02             & 11.19          & 0.02             \\
star 11 & 4395399303719163904 & 13.172         & 11.35          & 0.02           & 10.70          & 0.03           & 10.55          & 0.03             & 10.43          & 0.02             & 10.47          & 0.02             \\
star 12 & 1598160152636141568 & 10.780         & 9.10           & 0.02           & 8.53           & 0.03           & 8.37           & 0.02             & 8.32           & 0.02             & 8.35           & 0.02             \\
star 13 & 4535258625890434944 & 13.126         & 11.60          & 0.02           & 11.11          & 0.03           & 11.02          & 0.02             & 10.96          & 0.02             & 11.00          & 0.02             \\
star 14 & 2159020415489897088 & 12.505         & 10.81          & 0.02           & 10.21          & 0.02           & 10.06          & 0.02             & 10.01          & 0.02             & 10.07          & 0.02             \\
star 15 & 3713862039077776256 & 11.170         & 9.48           & 0.03           & 8.86           & 0.02           & 8.75           & 0.02             & 8.66           & 0.02             & 8.74           & 0.02             \\
star 16 & 4531308286776328832 & 11.832         & 10.12          & 0.02           & 9.56           & 0.02           & 9.45           & 0.02             & 9.36           & 0.02             & 9.37           & 0.02   \\ \hline         
\end{tabular}%
}
\end{table*}

\section{Atmospheric Parameters}
\label{atmospheric_section}
One of the primary goals of the work is to measure the atmospheric properties (i.e., effective temperature, surface gravity, metallicity, microturbulence) of these HVS candidates. High quality measurements of these properties are necessary to infer chemical abundances, and thus constrain the origins and production mechanisms of these fast stars.  Our spectra for the HVS candidates are low to mid SNR and appear metal poor based on visual inspection of the spectra. These data properties make a purely spectroscopic analysis challenging, as low SNR limits the number of weak absorption features we can use, and the metal poor nature implies that we must be careful about non-local thermodynamic equilibrium (NLTE) effects. To achieve the highest quality atmospheric parameters, we develop a workflow that combines spectroscopic, astrometric and photometric information simultaneously to find a self consistent model for the star similar to Section 3 of \cite{Henrique_2022}; however, we choose to use a spectral synthesis approach rather than a line-by-line synthesis used in the aforementioned study due to the low SNR of our spectra. Our spectroscopic analysis is done using methods and models which assume local thermodynamic equilibrium (LTE), departures from these assumptions can arise in the metal poor regime and may be substantial \citep[see e.g.,][]{Frebel_2013}, however the photometric information is less affected by this \citep[see e.g.,][and references therein]{Frebel_2013}. The photometric data also bypasses the problems of low SNR spectra, while being sensitive to both the effective temperature (\teff), and surface gravity (\logg). However, the photometric metallicity ($\mathrm{[Fe/H]}$) signal is weaker and heavily reliant on blue bands and extinction estimates. On the other hand, the spectroscopic fitting is more sensitive to the metallicity and microturbulent velocity ($\xi$) while largely agnostic about the presence of extinction. Hence, we measure \teff/\logg \ using photometry and metallicity/$\xi$ from spectroscopy. We employ python code \texttt{LoneStar} for the spectroscopic analysis (see Section \ref{sec:lonestar} for details) and the python package \texttt{Isochrones}\footnote{https://github.com/timothydmorton/isochrones} for the photometric analysis (see for Section \ref{sec:isochrones} details). 

The step-by-step fitting process is as follows:
\begin{enumerate}
    \item Fit the spectrum with \texttt{LoneStar} to find initial guesses for all atmospheric parameters (i.e., \teff, \logg, \feh, $\xi$)
    \item Fit the photometric data listed in Table \ref{tab:photometry} with \texttt{Isochrones} using values from LoneStar as a guess
    \item Re-Fit the spectrum using \texttt{LoneStar} holding \teff/\logg \ fixed from the photometric fit in step 2. A guess for $\xi$ is created using the surface gravity relationship from \cite{microturbulence_ref}, their Eq. 2. 
    \item Re-Fit the photometric data using \texttt{Isochrones} with the updated $\mathrm{[Fe/H]}$ from the previous step, allowing all parameters to vary
    \item repeat the previous two steps until the metallicity estimates from \texttt{LoneStar} and \texttt{Isochrones} converge\footnote{Convergence is defined as overlap in the 1$\sigma$ total error intervals for the metallicity estimates from \texttt{LoneStar} and \texttt{Isochrones}} or stability in the atmospheric parameters is attained
    
\end{enumerate}

Typically it takes a couple of iterations to reach termination (i.e., the metallicity is consistent or stable in both methods). Convergence in metallicity is preferable but not always achievable. Differences of up to $0.2$ in metallicity were found for some stars between the photometric and spectroscopic fits. This is in line with \cite{bochanski_18}, who find the mean spectroscopic and photometric metallicities of two clusters to be discrepant at the 0.15 dex level. Often this appeared with fits that had anomalously high extinction fits, using higher dust content to counteract higher metals. There are known shortcomings in photometric models of stars as well. In the event the two metallicity measurements do not agree within the total errors (i.e., the internal errors added in quadrature with the external errors) we use the spectroscopic metallicity. We reason that this represents the closest approximation to the real value because spectral lines are sensitive to the bulk abundance changes the metallicity represents. Metallicity and microturbulence are also not strongly correlated for those fits. In contrast, the photometric fits for metallicity show a strong degeneracy with extinction estimates even with strong priors on the dust because our blue band photometry does not place strong enough constraints on the isochrone fit. We found this discrepancy between the photometric and spectroscopic metallicity was also present for the test star we analyzed from \cite{Henrique_2022}, with a difference of $\sim0.15$ dex. However, if we consider only the spectroscopic metallicity, we find the same measurement as \cite{Henrique_2022}.

Internal errors for each parameter are derived from the method used to measure said parameter. \teff \ and \logg \ are measured using photometry and we use the posteriors from \texttt{Isochrones} as their internal uncertainties. $\xi$ is measured solely from spectroscopy. The internal error for $\xi$ from the posterior was small for all stars, and we took the largest value of 0.03 km~s$^{-1}$ as the assumed error for the entire sample. As discussed below, the external errors for $\xi$ are 2 orders of magnitude larger, so this choice does not materially change the results. Lastly, the metallicity is measured in both the photometric and spectroscopic approaches. We prefer and use the spectroscopic value because, as previously stated, we have more confidence in the accuracy of it. The internal error for the metallicity was taken as the quadrature sum of the internal errors from the photometric and spectroscopic posteriors. 

To evaluate the efficacy of our atmospheric parameter estimation, we compare our fits to the literature values for the standard stars. Since this study focuses on metal poor objects, we limit our comparison to objects with $\mathrm{[Fe/H]} \lesssim -0.5 \ \mathrm{dex}$. We find the following median offsets and dispersion $\Delta\mathrm{T_{eff}} = 181 \ \mathrm{K}, \sigma\mathrm{T_{eff}} = 40 \ \mathrm{K}, \Delta\log g = 0.07 \ \mathrm{dex}, \sigma\log g = 0.13 \ \mathrm{dex}, \Delta\mathrm{[Fe/H]} = 0.06 \ \mathrm{dex}, \sigma\mathrm{[Fe/H]} = 0.04, \Delta\xi =0.01 \ \mathrm{km \ s^{-1}}, \Delta\xi =0.28 \ \mathrm{km \ s^{-1}}$. The external error for each parameter is taken as the standard deviation of the difference, yielding  $\sigma_{\mathrm{ext}_{\mathrm{[Fe/H]}}} \sim 0.08 \ \mathrm{dex}$, $\sigma_{\mathrm{ext}_{\mathrm{T}_{\mathrm{eff}}}} \sim 40 \ \mathrm{K}$, $\sigma_{\mathrm{ext}_{\log g}} \sim 0.13 \ \mathrm{dex}$, and $\sigma_{\mathrm{ext}_{\xi}} \sim 0.28 \ \mathrm{km \ s^{-1}}$. For the literature comparison our sample included HD 122563, which is a Gaia benchmark star. We elected to use a microturbulence value of 1.8 $\mathrm{km \ s^{-1}}$ rather than the value of 1.13 $\mathrm{km \ s^{-1}}$ listed in \cite{benchmark_metal}. We calculate this revised value using the Gaia-ESO relationship. We prefer our revised value as the literature value seems very abnormal compared to even the paper it is listed in.

\subsection{LoneStar}
\label{sec:lonestar}

\texttt{LoneStar} is a python code written by T. Nelson to perform stellar atmospheric and abundance fitting for high resolution spectra. The goal of this package was to combine the benefits of traditional synthesis based approaches \citep[e.g., \texttt{BACCHUS}][]{bacchus2016} with a Bayesian framework to improve the error analysis and work at lower SNR. The code is organized into two modules, \texttt{abund} and \texttt{param}. The latter will be detailed here, with additional details for the abundance fitting provided in Section \ref{section:abund_methods}. 

The user designates an interpolator, a collection of wavelength regions of interest, which atmospheric parameters should be varied, and what priors to use for the Bayesian regression. The fitter then uses the Markov Chain Monte Carlo (MCMC) python package \texttt{emcee} \footnote{https://github.com/dfm/emcee} to maximize the posterior probability distribution. We typically require 18-24 walkers and around 3000 iterations to converge. We attempt to account for the following sources of error when minimizing the data: flux errors, interpolator reconstruction errors, and synthesis errors. To accomplish this, we introduce an error softening term for remaining unaccounted for terms to improve performance which is simultaneously fit along with the atmospheric parameters. The following parameters can be varied or fixed: \teff, \logg, \feh, $\xi$ and rotational broadening (\vsini). \vsini \ is applied on-the-fly using a convolution recipe from \cite{gray_book}, which assumes a limb darkening coefficient of $\epsilon = 0.6$. We use the wavelength sampling of ARCES for the atmospheric parameter fitting for a homogeneous analysis. This results in a downsampling of the data from TS by a factor of 2; however we have found this makes a negligible difference to the values fit for various test cases \citep[including all stars from][]{Nelson_2021}. Models are originally created with a wavelength sampling 3x higher than the TS data and subsequently downsampled to the ARCES wavelength space.

\texttt{The Payne} \citep{Payne} was used as the interpolator. We synthesized a library of $\sim11000$ spectra to train this artificial neural network with a single hidden layer containing 300 nodes\footnote{This structure differs from the one outlined in \cite{Payne} because we use 1 larger network for all pixels rather than a small network for each pixel in accordance with the current release of the Payne. This network architecture allows better modeling of pixel-to-pixel covariances and therefore should be more precise.}. To create our library of synthetic spectra we randomly sampled the following intervals: $3900 \ \mathrm{K} \le \mathrm{T_{eff}} \le 7000 \ \mathrm{K}$, $0 \ \mathrm{dex} \le \log g \le 5 \ \mathrm{dex}$, $-3  \le \mathrm{[Fe/H]} \le 1$. For each combination of \teff, \logg, and [Fe/H] we create three synthetic spectra by setting $\xi$ equal to 0, 1.5, and 2.6 \ $\mathrm{km \ s^{-1}}$. All synthetic spectra were constructed from MARCS model atmospheres \citep{MARCs} using \texttt{TURBOSPECTRUM} \citep{turbospectrum} for radiative transfer. MARCS models are calculated in 1D LTE. If the surface gravity is $\ge 3.0$ dex, plane-parallel models are used, and spherical models otherwise. If a combination of \teff, \logg, and \feh lies between MARCs models, an interpolation is done to create the specified model atmosphere. The atmospheric composition uses solar abundances from \cite{grevesse_solar_abunds} scaled by metallicity for most elements.  MARCs models use separate abundance estimates for C, N, and O \citep[see][Section 4 for more information]{MARCs}. We assumed the same composition for C, N and O as the models. We use Gaia-ESO line list version 5 \citep{Heiter2019} for atomic transitions. The line list includes hyperfine structure splitting for Sc I, V I, Mn I, Co I, Cu I, Ba II, Eu II, La II, Pr II, Nd II, Sm II. We also include molecular data for CH \citep{CH_data_14}, C$_2$, CN, OH, MgH (T. Masseron, private communication), SiH \citep{Kurucz_atomic_molecular_data}, TiO, FeH, and ZrO (B. Pelz private communication).

Wavelength masking is vital for an accurate atmospheric parameter fitting process because poorly modeled regions or problematic lines can alter the minimization. To begin, we limit the usable data to the range of $4500 - 6800$\ \AA. The limit on the blue side arises from a combination of reduced detector sensitivity, low source flux from our HVS candidates because we targeted late-type stars, and difficulties inherent to accurately placing the continuum in regions of dense metal absorption. This makes accurate continuum placement for metal rich stars challenging in the blue. Hence to be uniform in our treatment of program and standard stars we exclude data below 4500 \AA. The data showed wavelength calibration issues past 8000 \AA \ for some stars, therefore we excluded two lines at $\sim 8500$\ \AA \ from the line selection in \cite{Hawkins_2018}. With these two lines removed, the reddest line in our line selection for iron in the atmospheric parameter fitting was at $\sim6750$\ \AA, so an upper limit of 6800 \AA \ was used for the synthesis. Next we exclude features from ``bad'' pixels which can arise from the following: leftover cosmic rays\footnote{We also inflate errors nearby likely cosmic rays in case of bleeding between adjacent pixels. We use the \texttt{iSpec} cosmic ray detection function with a variation threshold of 0.15. For each index flagged as a cosmic ray, we inflate flux errors by a factor of 10 for the 10 closest pixels on the red side and the 10 closest on the blue side.}, scattered light features, or dead pixels. With this cleaned spectrum, we then mask wavelengths outside the vicinity of iron lines used in previous studies on metal poor stars by \cite{Hawkins_2018} and \cite{Ji_2020}. The line core is taken as the local minimum closest in wavelength to the line data. The extent of the wavelength window around each line is determined by a first derivative test, however adopting a small $\Delta\lambda$ window of 0.5 or 1 \AA\ around each iron line does not change the results. 




We use Bayesian regression to estimate the atmospheric parameters. Ordinary regression determines the best fit through minimizing the differences between the the error weighted sum of squared residuals (SSR) between the data and the model. Bayesian regression builds on this approach by including terms to represent the behavior of the model parameters based on previous knowledge. These additional terms are called priors. We initially adopt uninformative priors (i.e., uniform distributions) on all parameters. We limit the temperature to a range of 4000 to 6500 K based on the spectral types of the program stars. On subsequent iterations, where we fix \teff \ and \logg, we adopt Gaussian priors for [Fe/H] and $\xi$. The mean for [Fe/H] is taken as the output from \texttt{Isochrones}. The mean for $\xi$ is determined by inputting the \texttt{Isochrones} surface gravity estimate into the \cite{microturbulence_ref} relationship. We adopt standard deviations of 0.1 dex and 0.3 $\mathrm{km \ s^{-1}}$ for \feh \ and $\xi$, \ respectively. This choice represents our increased confidence in values of the parameters without being overly restrictive.


Upon finishing a fit, Lonestar writes the chain file for the MCMC, a small record of the parameter fits (including fixed and freed quantities), and some diagnostic plots to visualize how the fit performed. The best fit is the median. The upper and lower 1$\sigma$ errors are the 84th and 16th percentiles, respectively.

\subsection{Isochrones}
\label{sec:isochrones}
\texttt{Isochrones} is a package to fit MESA Isochrones and Stellar Tracks \citep[MIST,][]{Dotter_2016} models using the MultiNest wrapper PyMultinest \citep{Pymultinest} to photometric data. \texttt{Isochrones} also uses Bayesian regression for data fitting, so the user can specify initial parameter values and priors for those values. If priors are not specified, \texttt{Isochrones} adopts default distributions, we refer the interested reader to their package documentation for these.

Following the general procedure from \cite{Henrique_2022}, we input the following data: atmospheric parameters (\teff, \logg, [Fe/H]), the median photo-geometric distance estimates from \cite{Bailer-Jones_2021}, extinction estimates from the \texttt{dustmaps} python wrapper for \texttt{BAYESTARS} \citep{Green_2019}, and the photometric data for our HVS candidates described in Section \ref{external_data}. The dustmaps provided by \texttt{BAYESTARS} are 3D if the stars are inside the modeled volume. In cases where the star resides outside the modeled volume a 2D dustmap which integrates the modeled dustmap is used instead. All of our input quantities require error estimates. For the atmospheric parameters, we adopt 100 K, 0.5 dex, and 0.1 for \teff, \logg, and [Fe/H], respectively. We assume an error floor of 0.01 mag for $\sigma_{\mathrm{A_v}}$ computed by \texttt{BAYESTARS}. We adopt an error floor of 5 mmag for the photometry because it improved the fitting performance, similar to \cite{Henrique_2022}\footnote{ \cite{Clark_2022} finds an even higher error floor for photometry of 50 mmags is needed in their work with \texttt{Isochrones}.}.

We adopt the default priors for all quantities aside from metallicity and distance. For metallicity, we use a uniform prior between -4 and 0.5. For distance, we use a Gaussian prior centered on the median photo-geometric distance from \cite{Bailer-Jones_2021} and a standard deviation which is the difference in the upper and lower $1\sigma$ errors divided by two. We restrict the extinction to a range of 0 to $\mathrm{A_v} + \sigma_{\mathrm{A_v}} + 0.1$, where $\mathrm{A_v}$ is the estimate produced from \texttt{BAYESTARS}, $\sigma_{\mathrm{A_v}}$ is the 1$\sigma$ error estimate from \texttt{BAYESTARS}. The extinction in \texttt{Isochrones} is largely constrained by blue band photometry, however most of our HVS candidates lacked good photometry in the blue. Hence constraining the extinction to realistic values was necessary. In the absence of these tight constraints, the dust can deviate substantially from the \texttt{dustmaps} estimates. This deviation could be caused by imperfect models or data problems, where the dust value could compensate for these shortcomings. We note that the uncertainties derived from \texttt{Isochrones} do not include any systematics. The fitting process only uses one set of models and the uncertainties reported are solely the posteriors from the Bayesian distributions.

\section{Abundances}
\label{section:abund_methods}
Once the atmospheric parameters are determined, we measure abundances for up to 22 elements with the \texttt{abund} module of LoneStar. The following elements are measured\footnote{We attempted to measure Li. The only star which had a detection of Li was the contaminate Gaia DR3 source~id 4150939038071816320, which was a dwarf. The remainder of our sample were giants.}: Na, Mg, Al, Si, Ca, Sc, Ti, V, Cr, Mn, Fe, Co, Ni, Cu, Zn, Sr, Y, Zr, Ba, La, Nd, and Eu. This list includes members from the light/Odd-Z, Neutron Capture, $\alpha$ elements, and Fe-peak nucleosynthetic families. 

The \texttt{LoneStar} \texttt{abund} module synthesizes spectra at different [X/H] ratios. We synthesize spectra at $\mathrm{[X/H]} = 0, \pm 0.3, \pm 0.6$ in this fitting process. A model atmosphere is created using the best fit values determined from atmospheric parameter fitting. Synthetic spectra are created in the same manner as the atmospheric parameters with two exceptions. First, the spectra will have the abundance of the element of interest altered. Secondly, the spectra are created using a radiative transfer code rather than interpolation from a precomputed grid. During the initial abundance fitting, we do not assume an $\alpha$ \ enhancement based on the metallicity in order to be agnostic to the origins of these stars. The abundances change negligibly ($\mathrm{median}(\Delta\mathrm{[X/H]}) < 0.02$) when the average $\alpha$ abundances (i.e., Mg, Si, Ca, Ti\footnote{Ti is included here because an $\alpha$ enhancement in the MARCs models will include Ti.}) from the first iteration are fed into the analysis.

To measure the abundance for the species of interest, the user provides a line selection. The exact wavelength of the theoretical and observed line will primarily differ from imperfect wavelength calibration and other data reduction artifacts. Such discrepancies can be significant if uncorrected \citep[see e.g.,][Section 4.1]{jofre_benchmark_abunds}. We use the same line core and wing search algorithm as described for the Fe lines in Section \ref{sec:lonestar}. Then abundances for individual lines are found through $\chi^2$ minimization between the observation and synthetic spectra\footnote{If no local minimum is found using the input range of [X/H], the abundance range is adjusted to be centered around the abundance with the smallest $\chi^2$ in the test value set and the fit is repeated. The smallest $\chi^2$ value may occur on the upper or lower side of the abundance range. This process repeats up to 5 times, after which we conclude we are unable to adequately model the observation.}. We estimate the $1\sigma$ uncertainties using the width of the $\chi^2$ curve \citep[see e.g.,][]{fisher_stuff}. We neither downsample nor mask pixels in this step. For each line, plots of the data and synthesis are provided for visual inspection of the fit quality. During the fit process, a line may be rejected for lack of sensitivity over the [X/H] range used (i.e., no change in the $\chi^2$ values), the automatic windowing failing, inadequate sampling of the line in the data based on the window limits, and a few other pathologies. If a line is rejected based on this automatic assessment, the line data and cause of the rejection are recorded in a tracker object. These are saved for the user to review later. For any line fit, a quality flag is created indicating if there are problems with the fit (e.g., a reduced $\chi^2$ greater than 3 or less than 0.5). Once all lines for a species are either fit or rejected, the abundances and quality flags are tabulated and output for the user. The line list selection for all elements and all stars is given in Table~\ref{tab:line_subset}. We use a different line selection for metal poor stars (taken as $\mathrm{[Fe/H] < -0.5}$) and metal rich stars. This is primarily a caution for potential NLTE effects. In addition to modeling concerns, some lines may become measurable in the absence of dense absorption caused by higher metallicities (e.g., towards the blue end of the spectrum).

We use internal quality cuts to help filter out problematic abundance measurements from specific absorption features (e.g., Co from star 6 due to noise). These quality cuts will vary on a line-by-line and star-by-star basis therefore the final line selection for each star may be slightly different. We require all absorption lines used for abundance determination to be at least 3$\sigma$ detection, where we use a local SNR estimate with the relation from \cite{cayrel_eqw} to approximate the uncertainty in the equivalent width based on the continuum placement. We supplement our automatic quality flagging with visual inspection of all lines in our selection for 7 stars of varying atmospheric parameters and SNR.

Abundances reported are taken as the median of the lines that pass quality controls. The internal errors are estimated as the standard error (i.e., std(abundance)/$\sqrt{\mathrm{N_{lines}}}$). If only one line is present we take the uncertainty on $\chi^2$ as the internal error. To propagate the uncertainties from the atmospheric parameters we employ a sensitivity analysis in similar fashion to \cite{Hawkins_binary} and \cite{Nelson_2021}. For each parameter, we perturb the best fit model and derive abundances for this perturbed model atmosphere. The difference between the abundances from the best fit and perturbed model is the error introduced from that parameter. These abundance errors are added in quadrature with the line-by-line statistical errors for [X/H] to determine the total error for an abundance measurement. One limitation of this process is that it does not account for covariances in uncertainties between the atmospheric parameters. 


 The Fe line selection between the atmospheric and abundance fitting is different. For the atmospheric parameters, we use the union of lines from \cite{Hawkins_2018} and \cite{Ji_2020} whereas the abundances only use lines from the former. The change in line selection comes from distinct goals in the \texttt{param} and \texttt{abund} analysis. The former was tasked with creating a starting point so casting a wide net was desirable. The latter was a refinement of this fitting process and so we decided to use the line list the author was more familiar with. This amounts to $\sim 70$ fewer lines being used for Fe in the abundance determination compared to the metallicity fit. This change, along with quality selection cuts, produces an offset between the metallicity and iron abundance of $-0.03 \pm 0.07$ for the entire sample and $-0.01 \pm 0.07 $ if we only consider stars with metallicity below $-0.5$.

NLTE corrections for Ca \citep{NLTE_Ca}, Co \citep{NLTE_Co}, Fe \citep{NLTE_Fe_Ti, NLTE_Fe}, Mg \citep{NLTE_Mg_1, NLTE_Mg_2}, Mn \citep{NLTE_Mn}, Si \citep{NLTE_Si}, and Ti \citep{NLTE_Ti_1, NLTE_Fe_Ti} are accounted for on a line-by-line basis using online tables from \href{https://nlte.mpia.de/gui-siuAC_secE.php}{MPIA}. Star 5 (Gaia DR3 source~id 1396963577886583296), lies outside the atmospheric parameter range of these published values, therefore we do not attempt to apply a correction for this star.

\begin{table}
    \centering
    \caption{A portion of our line selection for each element, its atomic properties and the absolute abundance we derive for each absorption feature. A full machine-readable version, including abundances for each star for each line, is available online. The lines used will vary between stars because of the quality checks. $\chi$ is the excitation potential in eV, $\log gf$ is the logarithm of the oscillator strength $f$ multiplied by its statistical weight $g$, and log($\epsilon$) is the absolute abundance (after subtracting the Solar abundances) derived for this line. The solar abundances adopted are from \protect\cite{grevesse_solar_abunds}, except where described otherwise in Section \ref{section:abund_methods}. \label{tab:line_subset}}
    \begin{tabular}{cccccc}
    \hline
    \hline
         Identifier & Element & $\lambda$ & $\log gf$ &$\chi$ & $\log(\epsilon)$ \\
& & (\AA) & (dex) & (eV) & (dex) \\
\hline
star 7 & Cr I & 5247.56 & -1.59  & 0.961 & -7.96 \\
star 7 & Cr I & 5272.0 & -0.42  & 3.449 & -7.73 \\
star 7 & Cr I & 5296.69 & -1.36  & 0.983 & -7.87 \\
star 7 & Cr I & 5300.74 & -2.0  & 0.983 & -7.88 \\
star 7 & Cr I & 5304.18 & -0.67  & 3.464 & -7.30 \\
star 7 & Cr I & 5345.79 & -0.95  & 1.004 & -7.99 \\
star 7 & Cr I & 5348.31 & -1.21 & 1.004 & -8.02 \\
star 7 & Cr I & 5409.78 & -0.67 & 1.03 & -8.00 \\
star 7 & Cr I & 5628.64 & -0.74  & 3.422 & -7.23 \\
star 7 & Mn I & 4783.42 & -0.499 & 2.298 & -8.30 \\
\hline
    \end{tabular}

\end{table}

\section{Dynamical Analysis}
\label{kinematic_section}
We employ a dynamic analysis to assess whether these HVS candidates to answer two questions: 1) Which, if any, of the HVS candidates are unbound or marginally bound? 2) For the unbound or marginally bound objects, what systems might be progenitors for these fast moving stars? 

We use the python package \texttt{galpy}\footnote{https://github.com/jobovy/galpy} for this analysis. For each orbit, we used the radial velocities from our observations, the \cite{Bailer-Jones_2021} photo-geometric distances, with the remaining astrometry from Gaia DR3. In general, there was very good agreement between \cite{Bailer-Jones_2021} distances and those fit from \texttt{Isochrones}. To construct our covariance matrix, $\bold\Sigma$, for uncertainty analysis, we use the uncertainties and covariances for the right ascension (RA), declination (Dec), proper motion in RA (pmra), and proper motion in Dec (pmdec) from Gaia, and assume the RV and \cite{Bailer-Jones_2021} distances are uncorrelated. We propagate measurement uncertainties to our orbit integration and other derived kinematic quantities through Monte Carlo sampling of 
the multivariate normal distribution $N(\vec{\mu}, \bold\Sigma)$, where $\vec{\mu}$ is the measured value for each quantity. This sampling is repeated 1000 times.

We use the MWPotential2014 \citep{Galpy} to approximate the Galactic potential. This potential uses a Navarro-Frenk-White halo with a scale length of 16 kpc \citep{Navarro_1996}. A Miyamoto-Nagai potential \citep{Miyamoto_1975} with radial scale length of 3 kpc and vertical scale height of 280 pc is used for the disc. Finally, the bulge has a power-law density profile with an exponent of -1.8 and is exponentially tapered at 1.9 kpc. We assume current values for the solar position and kinematics as $R_0 = 8.122 \ \mathrm{kpc}, z_0 = 20.8 \ \mathrm{pc}$ \citep{r_sun_gc,z_sun_est}, and a solar motion of $(U_{\odot}, V_{\odot}, W_{\odot}) = (12.9, 245.6, 7.78)$ $\mathrm{km \ s^{-1}}$ \citep{v_x_sun,v_y_sun,v_z_sun}.

\section{Results}
\label{sec:results}

\subsection{Kinematics}
\begin{figure}
    \centering
    \includegraphics[width=1\columnwidth]{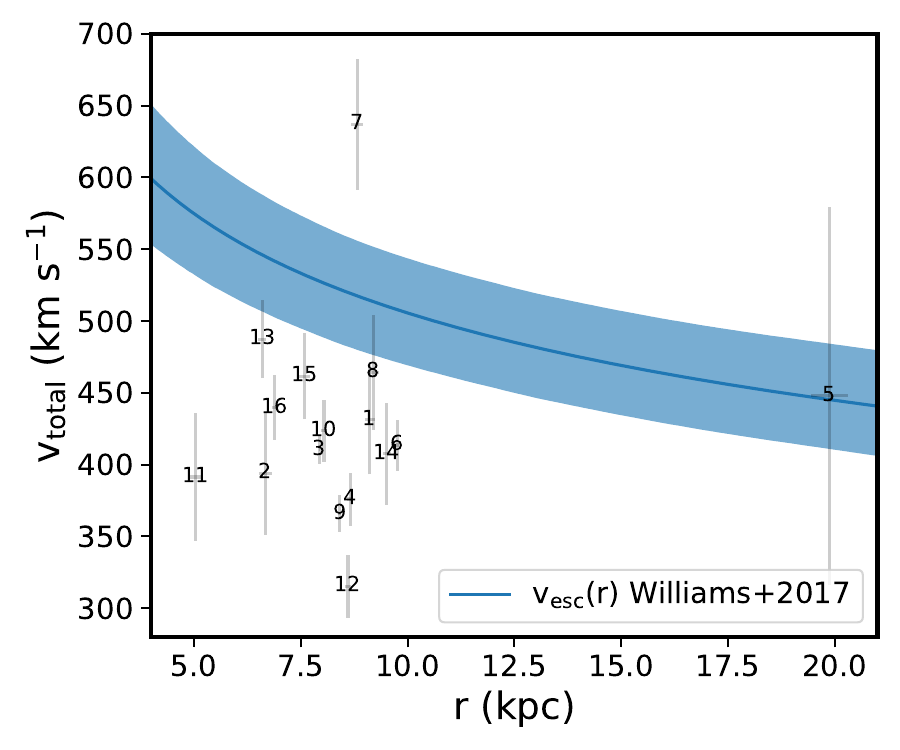}
    \caption{Displayed is the current spherical position and total velocity for each star in our sample. The numbers for each star correspond to their alias from Table \ref{tab:obs}. The error bars show the propagated uncertainties from the Monte Carlo sampling of the astrometry, RV, and distance uncertainties. The dark blue line represents the median escape velocity assuming the spherical model from \protect\cite{Williams_2017}, with the lighter blue contours corresponding to the 1$\sigma$ range, propagating the uncertainties in parameters from \protect\cite{Williams_2017}. Only star 7 is definitively unbound, star 5 is marginally bound, and star 8 could be unbound based on the overlap in the error bars.} 
    \label{fig:vesc}
\end{figure}

The kinematics of the candidate HVSs is used to determine whether these objects are gravitationally bound or unbound to the Galaxy, as well as where these stars may have been produced. This production location in turn constrains how these stars were accelerated. To access whether these candidate HVS are bound, we use their present day kinematics along with a model of the Milky Way's gravitational potential from \cite{Williams_2017}. We note the Milky Way model used in \cite{Williams_2017} differs from that used in Section \ref{kinematic_section}. In Figure \ref{fig:vesc} we show total velocity ($v_{\mathrm{total}}$) as a function of spherical distance from the GC (r), for our candidate HVS stars (labeled by their alias) and a Milky Way escape velocity curve with 1$\sigma$ \ uncertainties based on the model and uncertainties from \cite{Williams_2017}. We calculate $v_{\mathrm{total}}$ \ and r using the photo-geometric distances from \cite{Bailer-Jones_2021}, our ground-based RV measurements, and the remaining astrometry from Gaia DR3. The uncertainty band on the \cite{Williams_2017} model is created using Monte Carlo sampling of their model parameter uncertainties. From this work, we see that only star is likely unbound from the Galaxy, with stars 5, and 8 being marginally unbound (1$\sigma$ level). We have marked these stars in red in subsequent chemical plots to aid with their identification. 

We used catalogs of globular clusters and Milky Way satellites in \texttt{galpy} to determine if there were any clear candidate progenitors for stars 5, 7, and 8. These catalogs for globular clusters and Milky Way satellites are based on \cite{galpy_mw_gc} and \cite{galpy_mw_sat}, respectively. We integrated these systems using a similar framework as the previous section; however, we only integrated back 300 Myr. A star traveling with $100 \ \mathrm{km \ s^{-1}}$ in the radial direction would cover a distance of 30 kpc in this period, well outside the distances we expect our HVS to have traveled either from the outer Galaxy inward or vice versa. This choice also helps minimize potential inaccuracies from the uncertainty in the input phase space parameters $(x,y,z,v_x, v_y, v_z)$ \ and the Galactic potential. For all systems examined, the point of closest approach for our objects is at least 10 times the the half light radii of the candidate origin system. Doubling the integration length to 600 Myr, does not change the results. Extending the integration to 1.5 Gyr, the closest approach for stars 7 and 8 is $~\sim$ 1 kpc from the star systems examined. Interestingly, stars 7 and 8 share the same system of closest approach in their obits (NGC 6205), and the same second closest system (NGC 6341). Star 5 fairs worse, with the closest approach being Draco II at 3.5 kpc, and the second closest system being NGC 6229 at $\sim$ 8 kpc. 

We conducted a second round of kinematic analysis using a modified potential. \cite{bland_hawthorn_2016} estimate a dark matter halo mass \~50\% larger than the one used by default in MWPotential2014. In addition, \texttt{galpy} is capable of modeling the impact of the LMC's gravitational potential. \texttt{galpy} also provides a built in way to estimate the escape velocity from different symmetric potentials. Due to the LMC breaking cylindrical symmetry, we could only find an escape velocity estimate using the heavier dark matter halo potential. We find the escape velocity curve is unchanged from the \cite{Williams_2017} model used above. The top two systems change for Star 8 and are unchanged for stars 5 and 7. The system of closest approach for star 8 is NGC 5897, at a distance of $\sim$ 280 pc roughly $\sim$ 1.25 Gyo.

\subsection{Stellar parameters and abundances}
Atmospheric stellar parameters and chemical abundance measurements are displayed in Table \ref{tab:param_and_abund}. The full table includes both LTE and and NLTE corrected measurements when applicable.

\begin{table*}
\centering
\caption{A portion of the atmospheric parameters and abundances for the HVS candidates. Stars are labeled by their alias from Table \ref{tab:obs}. The $\mathrm{[Fe/H]}$ value is the abundance determined for iron rather than the metallicity from the atmospheric parameter fitting; however, these values are nearly identical ($|\Delta|<0.03$). $\xi$ is the microturbulence velocity. We only provide internal errors for \teff \ and \logg \ as the external errors are provided in the text and identical for each star. The internal errors for $\xi$ (i.e., $\sigma_{\xi}$) were negligible compared to the external error therefore we do not list them. Abundances which lacked measurements are given a nan. The error in abundance measurement is the total error described in Section \ref{section:abund_methods}. [X/Fe] values are provided in the digital version of this table. A full machine readable version of the table is available online.  \label{tab:param_and_abund}}
\resizebox{\textwidth}{!}{%
\begin{tabular}{cccccccccccccccc}
\hline
\hline
Alias   & $\mathrm{T_{eff}}$ & $\sigma_{\mathrm{T_{eff}}}$ & $\log g$ & $\sigma_{\log g}$ & $\mathrm{[Fe/H]}$ & $\sigma_{\mathrm{[Fe/H]}}$ & $\xi$         & $\mathrm{[Na/H]}$ & $\sigma_{\mathrm{[Na/H]}}$ & $\mathrm{[Mg/H]}$ & $\sigma_{\mathrm{[Mg/H]}}$ & $\mathrm{[Al/H]}$ & $\sigma_{\mathrm{[Al/H]}}$ & $\mathrm{[Si/H]}$ & $\sigma_{\mathrm{[Si/H]}}$ \\
        & (K)                & (K)                         & (dex)    & (dex)             & (dex)             & (dex)                      & (km s$^{-1}$) & (dex)             & (dex)                      & (dex)             & (dex)                      & (dex)             & (dex)                      & (dex)             & (dex)                      \\ 
        \hline
star 1  & 4654               & 39                          & 1.78     & 0.07              & -1.48             & 0.18                       & 1.88          & -1.45             & 0.24                       & -1.06             & 0.08                       & nan               & nan                        & -1.10             & 0.04                       \\
star 2  & 4588               & 30                          & 1.21     & 0.05              & -1.80             & 0.17                       & 1.87          & -1.82             & 0.12                       & -1.30             & 0.24                       & nan               & nan                        & -1.50             & 0.06                       \\
star 3  & 5158               & 31                          & 3.11     & 0.05              & -1.27             & 0.17                       & 1.76          & -1.29             & 0.24                       & -0.92             & 0.37                       & nan               & nan                        & -0.94             & 0.05                       \\
star 4  & 4763               & 20                          & 2.15     & 0.05              & -1.41             & 0.21                       & 1.82          & -1.20             & 0.23                       & -0.93             & 0.10                       & -1.00             & 0.09                       & -0.94             & 0.05                       \\
star 5  & 3987               & 11                          & 0.47     & 0.04              & -1.41             & 0.17                       & 1.97          & -1.18             & 0.08                       & -0.79             & 0.20                       & -0.85             & 0.08                       & -0.95             & 0.06                       \\
star 6  & 5352               & 56                          & 2.47     & 0.02              & -1.44             & 0.23                       & 1.88          & -1.16             & 0.04                       & -1.07             & 0.20                       & nan               & nan                        & -1.12             & 0.08                       \\
star 7  & 4817               & 13                          & 1.74     & 0.06              & -1.50             & 0.16                       & 1.82          & -1.34             & 0.09                       & -1.06             & 0.12                       & nan               & nan                        & -1.15             & 0.07                       \\
star 8  & 5099               & 47                          & 1.93     & 0.09              & -1.92             & 0.11                       & 1.83          & -1.25             & 0.20                       & -1.64             & 0.11                       & nan               & nan                        & nan               & nan                        \\
star 9  & 5601               & 41                          & 3.34     & 0.05              & -1.07             & 0.17                       & 1.78          & -0.69             & 0.10                       & -0.63             & 0.04                       & nan               & nan                        & -0.82             & 0.06                       \\
star 10 & 5025               & 56                          & 2.60     & 0.08              & -1.22             & 0.15                       & 1.97          & nan               & nan                        & -0.94             & 0.36                       & nan               & nan                        & -0.85             & 0.12                       \\
star 11 & 4469               & 11                          & 1.31     & 0.05              & -1.50             & 0.19                       & 1.95          & -1.34             & 0.16                       & -1.01             & 0.05                       & nan               & nan                        & -1.09             & 0.06                       \\
star 12 & 4526               & 16                          & 1.21     & 0.03              & -1.82             & 0.18                       & 1.88          & -1.80             & 0.07                       & -1.46             & 0.22                       & nan               & nan                        & -1.56             & 0.05                       \\
star 13 & 5048               & 25                          & 2.17     & 0.05              & -1.84             & 0.09                       & 2.36          & nan               & nan                        & nan               & nan                        & nan               & nan                        & nan               & nan                        \\
star 14 & 4689               & 19                          & 1.43     & 0.04              & -2.41             & 0.13                       & 1.64          & nan               & nan                        & -1.99             & 0.22                       & nan               & nan                        & nan               & nan                        \\
star 15 & 4485               & 50                          & 1.17     & 0.09              & -1.50             & 0.22                       & 1.87          & -1.30             & 0.09                       & -1.09             & 0.13                       & -1.07             & 0.09                       & -1.12             & 0.07                       \\
star 16 & 5023               & 97                          & 1.99     & 0.13              & -1.80             & 0.17                       & 1.87          & -1.78             & 0.08                       & -1.54             & 0.15                       & nan               & nan                        & -1.41             & 0.20   \\ \hline                   
\end{tabular}%
}

\end{table*}

\subsection{Comparison to Prior Works}
As part of our analysis, we observed stars from the Gaia benchmark stars \citep{benchmark_metal, benchmark_temperature_grav} and stars from \cite{Bensby_2014} to assess the differences in the atmospheric and abundance fits. We find a median offset of $\sim -181 \ \mathrm{K}$ in effective temperature. Our analysis of the stars from \cite{Hawkins_2018} show a similar offset of $\sim - 187 \ \mathrm{K}$ when compared with the previous spectroscopic temperatures. We suspect this offset arises from the discrepancy in photometric and spectroscopic temperatures. We offer two lines of evidence to support this hypothesis. First, in Table 3 from \cite{Hawkins_2018}, photometric temperatures from Gaia DR2 are provided, which show an offset of $\sim - 173 \ \mathrm{K}$. Second, Figure 2 from \cite{Frebel_2013} finds an offset of $\sim-200 \ \mathrm{K}$ between the photometric and initial spectroscopic temperatures (i.e., before NLTE considerations). Assuming we can apply the general relationship from very metal-poor objects to less metal-poor stars, this would explain the offset in temperature. Additional discussion on the impact of NLTE on effective temperature can be found in e.g., \cite{Korn2003} and \cite{Mucciarelli_2020}.

Offsets in the other atmospheric parameters are seen for the standard star sample. Comparing the atmospheric parameters for the stars in the \cite{Hawkins_2018} sample we see $\Delta \log g = -0.9/-0.52~\mathrm{dex}; \Delta \mathrm{[Fe/H]} = -0.44/-0.27~\mathrm{dex}; \Delta \xi = 0.1/-0.24~\mathrm{km~s^{-1}}$ for the \texttt{LoneStar} + \texttt{Isochrones} / \texttt{LoneStar} only fits respectively. These differences are an order of magnitude larger than those for the standard stars. These offsets could arise from differences in the treatment of NLTE, the low SNR of the data, and the fitting methods employed. \cite{Hawkins_2018} gauges the influence of NLTE effects by redoing their fits using the photometric temperature instead, finding offsets of up to 0.3 dex in metallicity. 

\begin{figure}
    \centering
    \includegraphics[width=1\columnwidth, trim={0 0.5cm 0 0},clip]{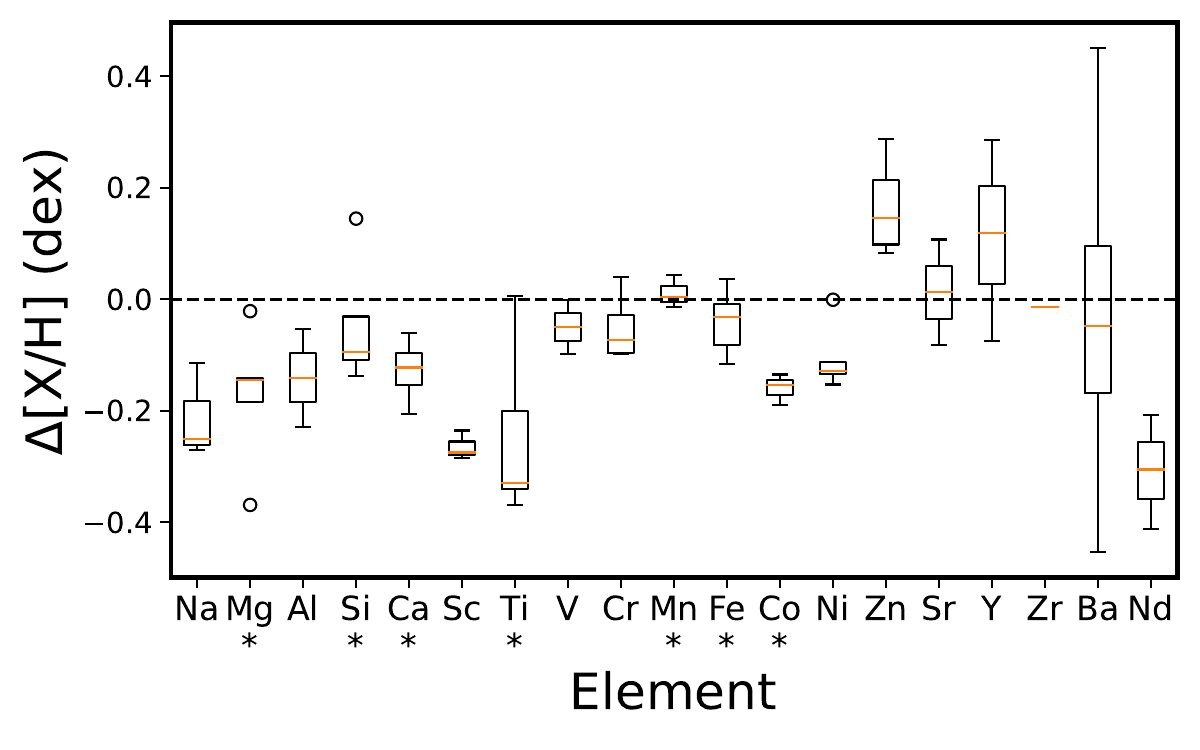}
    \caption{Differences in abundances derived for stars in \protect\cite{Bensby_2014, Battisini_2015, Battistini_2016} and this study. Elements which use NLTE corrections in this calibration plot are marked with asterisks below their label.} The disagreement between our measurements and the literature is reduced when we only examine LTE abundances and remove offsets caused by differences in metallicity measurements (i.e., use $\Delta\mathrm{[X/Fe]}$ instead of $\Delta\mathrm{[X/H]}$. Remaining differences are likely a consequence of differences in the \teff \ and $\xi$ \ parameters. We find differences of $\sim150$ \ K in \teff \ and $0.4\ \mathrm{km \ s^{-1}}$ in $\xi$.
    \label{fig:bensby_comparison}
\end{figure}

Abundance differences between \cite{Bensby_2014,Battisini_2015,Battistini_2016} and this study are shown in Figure \ref{fig:bensby_comparison}. Three elements lack literature comparisons: Copper (Cu), Lanthanum (La), and Europium (Eu). Copper and Lanthanum measurements were not available from the literature studies we referenced. No suitable measurements of Europium were found in our observations after filtering through our quality criteria.

There are several plausible sources for these abundance offsets; we will consider differences in atmospheric parameters and NLTE corrections. NLTE corrections do not have a consistent affect on the abundances. For Ti and Co, they increase the offset relative to a pure LTE comparison by $\sim0.1$ dex, while the rest have negligible changes (i.e., $\Delta < \pm 0.05$ dex). To gauge the significance of the atmospheric parameters, we use the stars from \cite{Hawkins_2018} as a proof of concept. After controlling for changes in bulk metallicity (i.e., using [X/Fe] rather than [X/H]) we find comparable offsets in Mg, Si, Ca, Zn, Sr, Nd, Y as observed in the data. Still further discrepancies could arise from the atomic data, the line selection, the visual inspection, continuum placement, and so on. 

When comparing our abundance measurements to the literature abundances, we do not use NLTE corrections from Section \ref{section:abund_methods} unless otherwise specified because several of the comparison studies \citep[e.g.,][]{Hawkins_2018} only compute LTE abundances. In addition, we do not rescale our data based on the reference stars from \cite{Bensby_2014, Battisini_2015, Battistini_2016} because this rescaling cannot be done uniformly for all the literature samples we compare our data with.

\subsection{Literature Sources}
For context in Figure \ref{fig:alpha_elements_combined} and Figure \ref{fig:combined_xfe}, we include measurements from studies on the thin and thick disc \citep{Bensby_2014, Battisini_2015, Battistini_2016}, the bulge \citep{bensby2010, gonzalez_2015}, the inner halo \citep{nissen_2010}, metal poor halo stars \citep{Yong_2013, Roederer_2014}, the LMC \citep{vanderswaelmen_2013}, and Fornax \citep{letarte_2010}. We have also included data from other studies on the chemistry of hyper/high velocity stars \citep{Hawkins_2018, Henrique_2022}. We are comparing LTE abundances to one another in these plots. 

\subsection{Chemical Abundances}
The goal of this work is to use chemical tagging \citep{Freeman2002}  to constrain the origins of our late-type HVS candidates. Chemical evolution of the interstellar medium (ISM) is a fundamental ingredient for chemical tagging because most stellar abundances reflect the composition of their progenitor ISM. Broadly, a generation of stars will form with some initial composition. As the stars age, their interior composition will change from fusion; however, the surface composition remains roughly constant over their lifespan and hence can act as a fossil record of the progenitor system. This modified stellar composition is then dispersed into the ISM through some flavour of supernova (type Ia, type II, etc.), stellar winds, or other mechanism (e.g., kilonova). The chemical evolution of the ISM depends on the availability of new materials (i.e., the amount and type of feedback) as well as the mixing efficiency of said materials with the extant gas. The type and timescales of the feedback are dependent on mass, and to a lesser extent metallicity of the stellar population \citep[see e.g.,][and references therein]{nomoto_2013}. 

For the first $\sim1 \ \mathrm{Gyr}$, massive stars are thought to be the primary contributor to the chemical evolution of the ISM owing to their relatively short lifetimes compared to low mass stars \citep[see e.g.,][Section 1.3]{gilmore_1989}. Hence at low metallicities (e.g., $\mathrm{[Fe/H]} \lesssim -1$ for the solar neighborhood), the abundance patterns of the Galaxy reflect the yields from massive stars. These yields have a metallicity dependence \citep[see e.g.,][]{nomoto_2013}. After this period, feedback from lower mass stars (e.g., AGB winds, type Ia supernova) becomes increasingly important as more low mass stars reach the point at which they can expel their matter into the surrounding environment. Since low mass stars are far more numerous than higher mass stars \citep[e.g.,][]{kroupa_2003}, eventually the feedback of materials into the ISM from the lower mass stars will tend to dominate the present day ISM composition in areas of continuous star formation within the Galaxy. The metallicity at which low mass stars start becoming important is dictated by the star formation rate. The total mass of star forming matter and the initial mass function (along with the metallicity distribution function) play a similarly pivotal role in what feedback mechanisms are possible to subsequently modify the ISM.

\subsection{$\alpha$ elements: Mg, Si, Ca, Ti}
\label{alpha_el_section}

\begin{figure*}
    \centering
    \includegraphics[width=2\columnwidth]{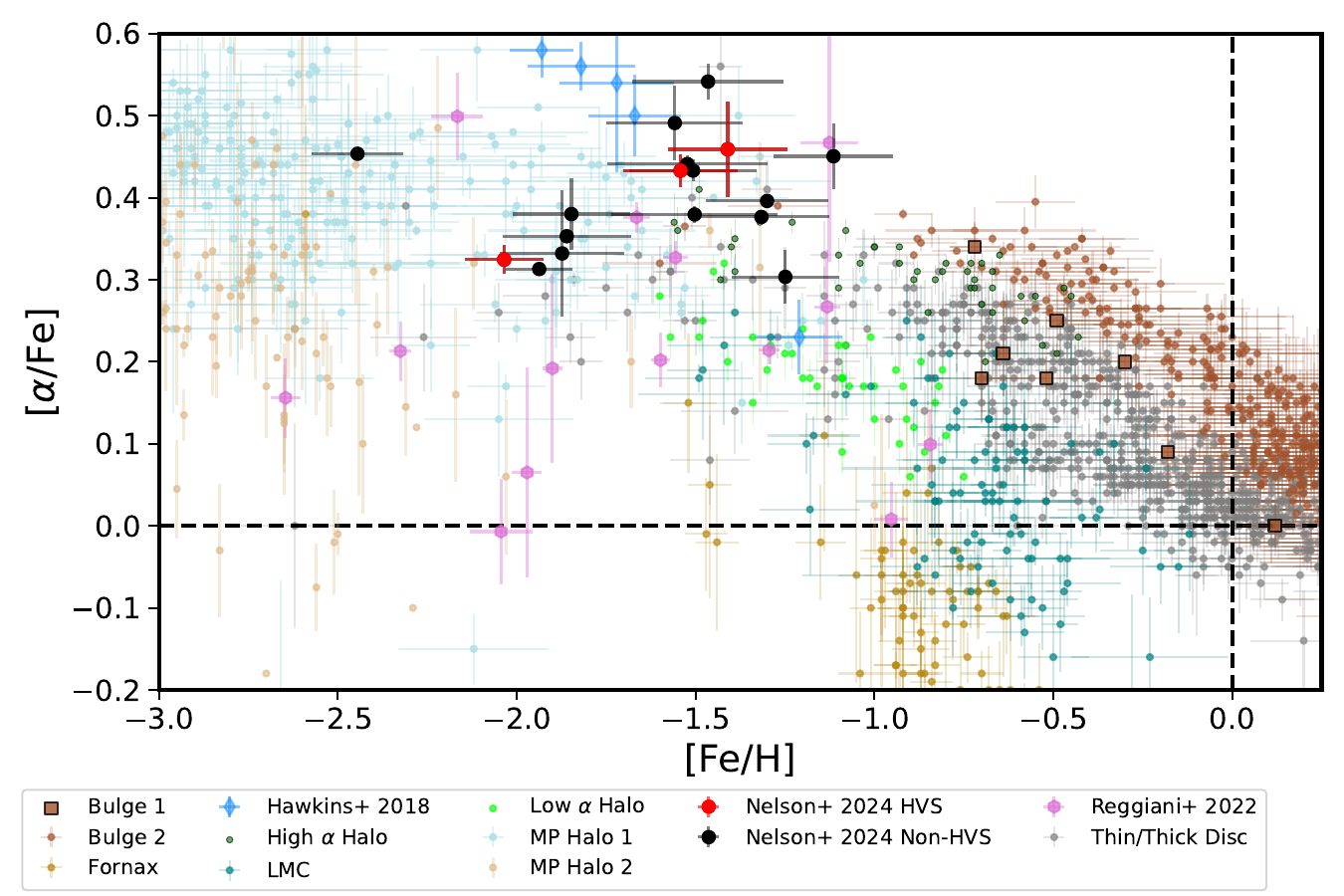}
    \caption{[$\alpha$/Fe] abundance measurements and errors as a function of metallicity are plotted. Data points from the HVS candidates found in Section \ref{kinematic_section} are shown in red, the remainder of the sample is shown in black. [$\alpha$/Fe] is taken as the median of [Mg/Fe], [Si/Fe], and [Ca/Fe]. If not all three elements have measurements, we take the average instead. All abundances shown are taken to be in LTE. For reference, in each panel we also show the abundance ratios of the thin and thick disc \protect\citep[][in grey]{Bensby_2014, Battisini_2015, Battistini_2016}, the high $\alpha$ \ halo \protect\citep[][in green]{nissen_2010}, the low $\alpha$ \ halo \protect\citep[][in bright green]{nissen_2010}, the metal poor halo \citep[][in light blue]{Roederer_2014} \protect\citep[][in tan]{Yong_2013}, and the the bulge \protect\citep[][in brown]{bensby2010, gonzalez_2015}. We include abundances from two contemporary studies on the abundances of hyper velocity candidates as well, \protect\cite{Hawkins_2018} in blue, and \protect\cite{Henrique_2022} in violet. Abundances for the LMC from \protect\cite{vanderswaelmen_2013} (teal) and Fornax from \protect\cite{letarte_2010} (gold) are also shown for additional context.}
    \label{fig:alpha_elements_combined}
\end{figure*}

\begin{figure*}
    \centering
    \includegraphics[width=2\columnwidth,trim={0 2.4cm 0 0},clip]{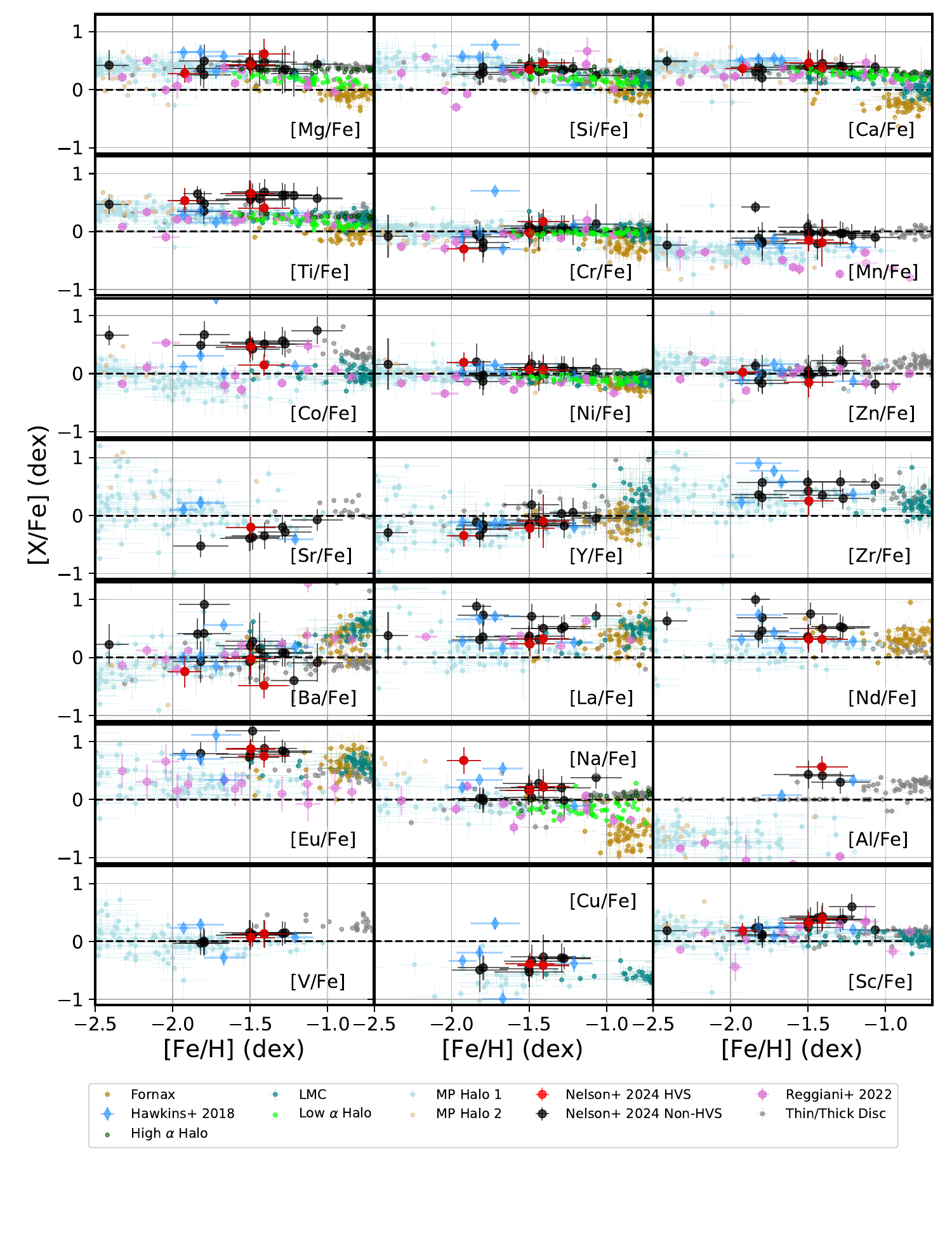}
    \caption{{[X/Fe] abundance measurements and errors as a function of metallicity are plotted. Data points from the HVS candidates found in Section \ref{kinematic_section} are shown in red, the remainder of the sample is shown in black. All measurements compared are in LTE.} For reference, in each panel we also show the abundance ratios of the thin and thick disc \protect\citep[][in grey]{Bensby_2014, Battisini_2015, Battistini_2016}, the high $\alpha$ \ halo \protect\citep[][in green]{nissen_2010}, the low $\alpha$ \ halo \protect\citep[][in bright green]{nissen_2010}, the metal poor halo \citep[][in light blue]{Roederer_2014} \protect\citep[][in tan]{Yong_2013}, and the the bulge \protect\citep[][in brown]{bensby2010, gonzalez_2015}. We include abundances from two contemporary studies on the abundances of hyper velocity candidates as well, \protect\cite{Hawkins_2018} in blue, and \protect\cite{Henrique_2022} in violet. Abundances for the LMC from \protect\cite{vanderswaelmen_2013} (teal) and Fornax from \protect\cite{letarte_2010} (gold) are also shown for additional context.}
    \label{fig:combined_xfe}
\end{figure*}

The $\alpha$ elements are formed through the consecutive addition of helium nuclei \citep[$\alpha$-particles, see e.g.,][]{alpha_start_paper}. Titanium, while not formed in the same pathway \citep[see e.g.,][]{Ti_production}, often follows the same trends so it is frequently included in this family of elements \citep[see e.g.,][]{Hawkins_stream}. The yields from core-collapse supernova (type II/Hypernova) dominate the dispersal of these elements and happen on shorter timescales than type Ia SNe. In the solar neighborhood, this manifests as an $[\alpha/\mathrm{Fe}]$ plateau of $\sim0.4$ for $\mathrm{[Fe/H]}\lesssim -1$. This can be seen in Figure \ref{fig:alpha_elements_combined}, where we have taken $\mathrm{[\alpha/Fe]}$ as the median of the [Mg/Fe], [Si/Fe], and [Ca/Fe] abundances measured for that star. The lower yield of $\alpha$ elements compared to iron (and Fe-peak elements) present in type Ia supernova causes  $[\alpha/\mathrm{Fe}]$ to decrease with increasing metallicity for $ -1 \lesssim \mathrm{[Fe/H]} \lesssim 0$ \citep[see e.g.,][]{lambert_1987, wheeler_1989, weinberg_2019}. At solar metallicities, $\mathrm{[\alpha/Fe]}\sim 0$. The inflection point at $\mathrm{[Fe/H]}\sim -1$ is referred to as the `knee'. The inner halo, bulge, thin disc, and thick disc all have distinct locations in the [$\alpha$/Fe] vs [Fe/H] plane, corresponding to their evolution; however, the boundaries between the regions are not always well defined \citep[see e.g.,][]{feltzing_2013, hawkins_2015_components}. Further, there are also signatures for accreted systems, which show a `knee' at metallicities lower than -1 dex.

Results for individual $\alpha$ \ elements are shown in Figure \ref{fig:combined_xfe}. We also show the combined abundance pattern in Figure \ref{fig:alpha_elements_combined}, where we have taken $\mathrm{[\alpha/Fe]}$ as the median of the [Mg/Fe], [Si/Fe], and [Ca/Fe] abundances measured for that star.

\begin{figure}
    \centering
    \includegraphics[width=1\columnwidth]{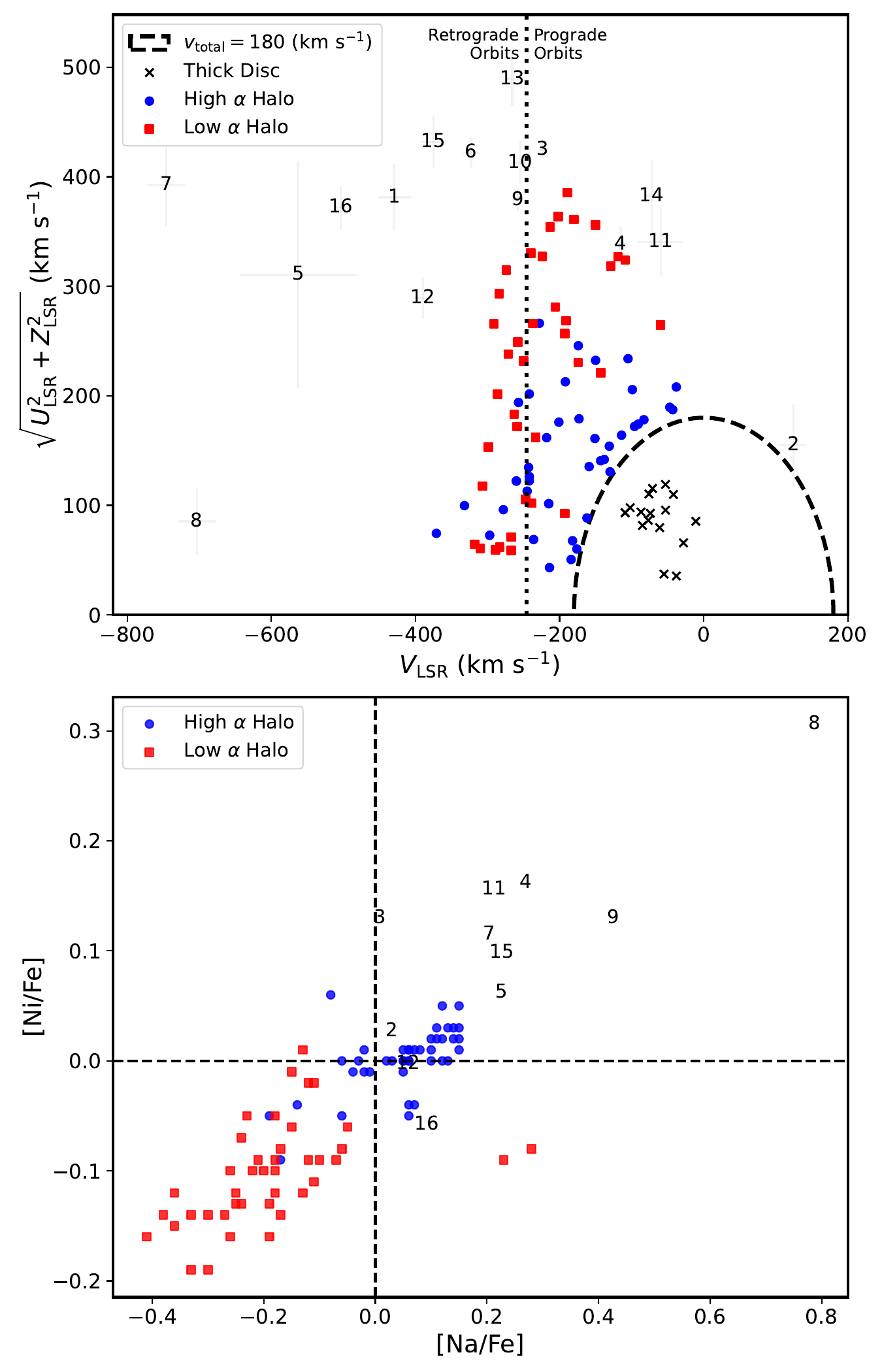}
    \caption{Top: A Toomre diagram for the HVS candidates in this paper. For ease of reference, each star in the paper is labeled by the number in its alias (i.e., star 7 is labeled 7). Data from \protect\cite{nissen_2010} is included for comparison. The high $\alpha$, low $\alpha$, and thick disc data from \protect\cite{nissen_2010} are labeled HA (blue circle), LA (red square), and TD respectively (black cross). Bottom: A comparison of the [Na/Fe] and [Ni/Fe] abundances for the candidate HVSs. Aliases are used for points similar to Figure~\ref{fig:toomre}. Data from \protect\cite{nissen_2010} is included for comparison and labeled as in Figure~\ref{fig:toomre}. The abundances used and compared to in this panel use LTE.}
    \label{fig:toomre}
\end{figure}

Overall, we find relatively good agreement between the chemical patterns of our stars and the inner stellar halo. This is consistent with the picture from \cite{Hawkins_2018}. However, we do not see a significant low $\alpha$ component in contrast to \cite{Henrique_2022, quispe_2022} which are both follow-up studies on the chemistry of HVS candidates. We examined all stars with $\mathrm{[\alpha/Fe]} \lesssim 0.3 $ to check if these objects were consistent with accreted origins and the so-called `low $\alpha$ \ halo' from \cite{nissen_2010}. \cite{nissen_2010} show that the low and high $\alpha$ \ halos cluster differently in Toomre space and in [Ni/Fe] vs [Na/Fe] space. The usefulness of the Toomre space clustering is hampered by our selection criteria of fast moving stars. In the Toomre space, shown in Figure~\ref{fig:toomre}, our entire sample of candidate HVSs appear more consistent with the fast moving low $\alpha$ halo compared to the slower high $\alpha$; however, most of our sample appears chemically consistent with the high $\alpha$ \ halo. This is also found when we compare the [Ni/Fe] vs [Na/Fe], shown in Figure~\ref{fig:toomre}, finding no stars which are unambiguously in the low $\alpha$ \ halo cluster.

\subsection{Light/Odd-Z elements: Na, Al, V, Cu, Sc}
\label{odd_z}
Odd-Z elements are produced in a variety of nucleosynthetic pathways. Sodium (Na) and Aluminum (Al) can both experience strong NLTE effects at low metallicities \citep[see e.g.,][and references therein]{kobayashi_2020}, making their interpretation difficult. Figure \ref{fig:combined_xfe} displays our measurements of Na, Al, V, Cu, and Sc in [X/Fe] vs [Fe/H] space (black circles). As before, we see that our stars are consistent with the stellar halo. Unlike \cite{Henrique_2022}, we do not see any evidence in [Na/Fe] of our stars being consistent with a dwarf or accreted galaxy. We find [Al/Fe] which over 1 dex greater than other literature we compared to. This difference in [Al/Fe] might be a result of different line selection between our study and \cite{Yong_2013, Roederer_2014, Henrique_2022}, all of which use lines close to $\sim 4000$ \AA, a section of data that we discarded during the reduction (see Section \ref{atmospheric_section}). We instead use 5557.06, 6696.02, 6698.67 \AA \ to measure the Aluminum abundance. These are weak lines in our program stars, so we are only able to measure Al in a handful of spectra. Applying NLTE corrections for Na does not meaningfully change the offsets between our values and \cite{Henrique_2022} for our line selection, so we conclude that most of our stars are likely not accreted or debris from a satellite galaxy.

\subsection{Fe-peak elements: Cr, Mn, Co, Ni, Zn}
The Fe-peak elements are primarily synthesized with Type Ia and core collapse supernovae \citep[e.g.,][]{nomoto_2013}. These elements largely trace the iron abundance.  As such, most of these elements are expected to have a roughly flat trend of [X/Fe] against metallicity. We plot our results of [Cr, Mn, Co, Ni, Zn/Fe] (black circles) in Figure \ref{fig:combined_xfe}. We find further evidence that these stars are likely from the halo based on their agreement with \cite{nissen_2010} and \cite{Hawkins_2018} in [Ni/Fe] and [Cr/Fe]. Our Cobalt (Co) abundances tend to be higher than \cite{Yong_2013, Roederer_2014} but are within the range seen by \cite{Hawkins_2018} and \cite{Henrique_2022}.

\subsection{Neutron Capture Elements: Sr, Y, Zr, Ba, La, Nd, Eu}
The neutron capture elements are commonly split into those which are primarily produced in the slow neutron capture process (s-process) and the rapid neutron capture process (r-process). The s-process elements (Sr, Y, Zr, Ba, La, and Nd) are formed in AGB stars and then returned to the ISM through stellar winds. By contrast, r-process elements are formed with rapid neutron capture. The exact nature of what processes drive this is still an open area of research, with binary neutron star mergers being one such candidate \citep{r_capture_production}. The neutron-capture element abundance ratios for our stars (black circles) for [Sr, Y, Zr, Ba, La, Nd, Eu/Fe] as a function of metallicity are shown in Figure \ref{fig:combined_xfe}. Of the neutron capture elements we measure, all are members of the s-process group except Eu. These elements are consistent with the stellar halo.



\section{Discussion}
\label{discussion}

In our sample, we find only one star that seems unbound based on the adopted escape velocity from \cite{Williams_2017}. Further, there is one marginally bound star and one star which could potentially be marginally bound on the overlap in its $v_{\mathrm{total}}$ uncertainties and the escape velocity uncertainties. 

\subsection{Possible origins for unbound or marginally bound stars 5, 7, and 8}



\subsubsection{Globular Clusters}
\begin{figure}
    \centering
    \includegraphics[width=1\columnwidth, trim={0 2cm 0 0}, clip]{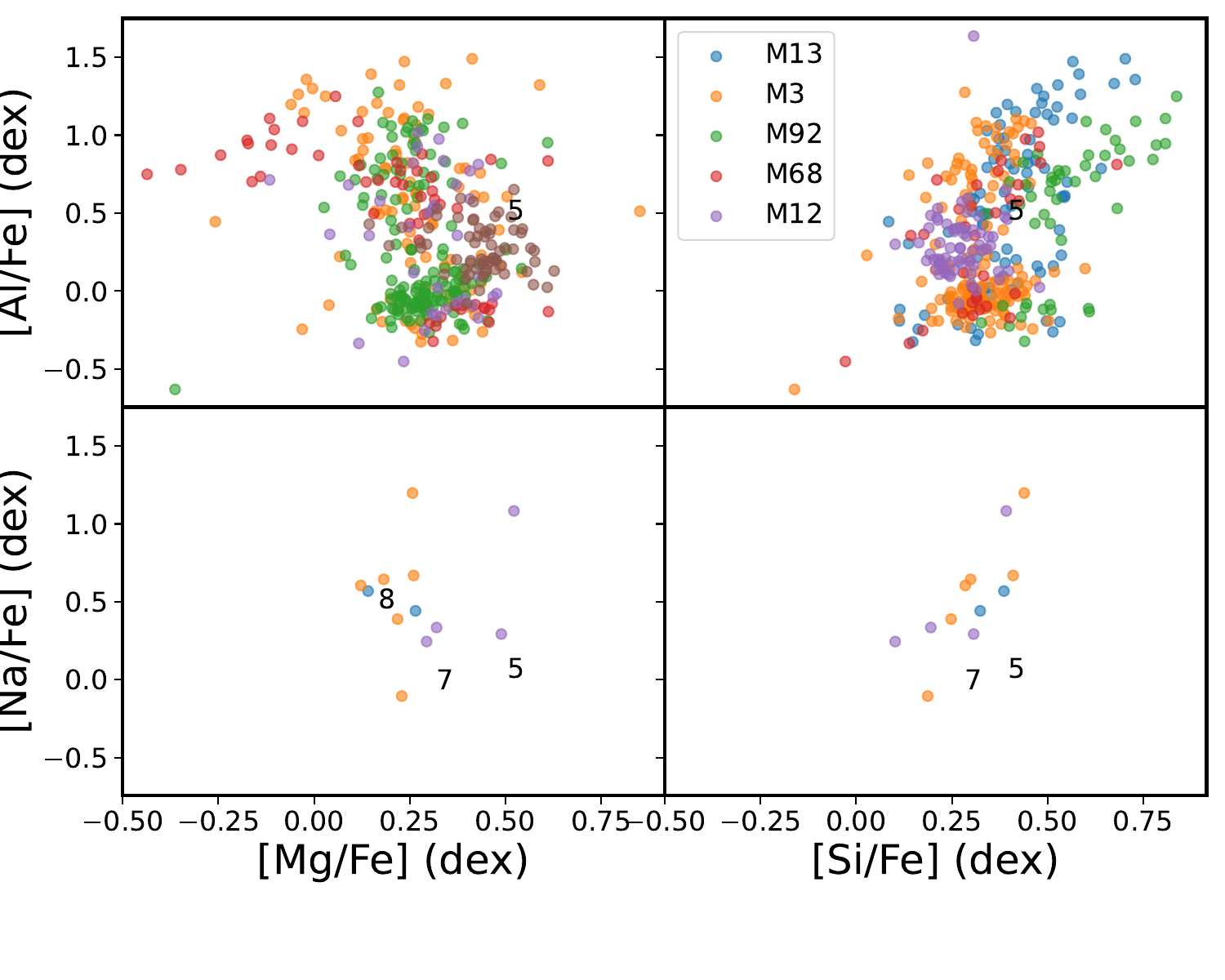}
    \caption{ Abundance ratios for the candidate HVSs compared to those known to be observed in globular clusters. Background data is taken from \protect\cite{Masseron_2019}. If the star lacked 1 of the measurements in a given panel it was excluded from that panel.} 
    \label{fig:globular_cluster}
\end{figure}

We plotted our stars against known abundance trends for globular clusters \citep[see e.g., Figure 2 in][]{Gratton_2019}. Due to SNR and weakness of the Al features we had available, we were only able to measure Al for star 5. All of the stars had abundances within the range seen for a few globular clusters studied in detail from \cite{Masseron_2019, Meszaros_2020} (M13, M3, M92, M68, and M12). We found that none of our stars were enhanced in Al while simultaneously being depleted in Mg, thus are not consistent with being second generation globular cluster stars. The abundance plots are shown in Figure \ref{fig:globular_cluster}.


\cite{Cabrera_2023} provides 50\% and 90\% credible phase space regions for stars dynamically ejected out of 148 globular clusters in the Milky Way. For stars 7 and 8 we find no suitable clusters. Star 5 is at a much greater distance and therefore the proper motion and sky locations are not as constraining on this star's origins.  However, based on the second kinematic analysis, it could be plausible for Star 8 to originate from NGC 5897. There is some probability based on the \cite{Cabrera_2023} models for star 8 to be located in is present region. Star 8 also has a chemical make up which agrees well with the previous characterization of NGC 5897 from \cite{NGC5897_chem}. \cite{NGC5897_chem} measure a metallicity of $\sim- 2.04$ and $\mathrm{\alpha/Fe}]\sim 0.34$ based on 7 stars from the cluster. We measure star 8 with a metallicity of $\sim- 1.97$ and $\mathrm{\alpha/Fe}]\sim 0.33$, which are within the measurement errors. This conclusion is hampered by our [Na/Fe] measurement which is $\sim0.7$ compared to their highest value of $\sim0.6$. They note that NLTE corrections at these parameter ranges might account for a difference of -0.05 dex. Higher SNR follow-up observations of star 8 would be useful to confirm this possible origin. These observations would also be useful for measuring more elements useful for studying populations of stars in globular clusters (e.g., O, Si, Al).

\subsubsection{LMC}
There is interest in detecting hyper velocity stars from the LMC, which could offer indirect evidence for the existence of a massive black hole in the center of the LMC \citep{lmc_hvs_smbh}. \cite{boubert_2016} provide on sky distributions for LMC stars ejected through the Hills' mechanism and \cite{boubert_2017} create phase space distributions for LMC run away stars. Our on sky distribution of stars is in the lowest or second lowest density contours for both scenarios. 

As pointed out in \cite{Henrique_2022}, due to the lack of LMC stars over the metallicity range covered by our objects, ruling out these stars came from the LMC based on chemistry alone is difficult. However, we see no positive evidence (i.e., the chemical patterns are not consist with LMC origins) favoring the LMC over the stellar halo based on chemical abundances. 

\subsubsection{Low $\alpha$ \ halo or Accreted system}
As discussed in Section \ref{alpha_el_section}, there are no stars in our sample that appeared to simultaneously satisfy the Toomre clustering, $\mathrm[\alpha/Fe] \lesssim 0.3$, and [Ni/Fe] vs [Na/Fe] grouping that \cite{nissen_2010} found for the low $\alpha$ \ halo. Star 8 appears to have a lower [$\alpha$/Fe] value but has very high [Na/Fe] and [Ni/Fe] making it incompatible with the \cite{nissen_2010} low $\alpha$ \ halo origin. Therefore we can rule out accretion as an origin for these HVSs. We can also rule out origins from a satellite galaxy with Fornax like chemistry due to the disagreements in [Na/Fe] seen in Figure \ref{fig:combined_xfe} and the [$\alpha$/Fe] seen in Figure \ref{fig:alpha_elements_combined}.

\subsubsection{Galactic Center}
\cite{boubert_2016} provide an on sky distribution for stars ejected from the GC.  Our stars fall outside the main density contours of this figure. This is supported by our orbit integration that finds stars 5, 7, and 8 have not originated within 7 kpc of the GC. 

\subsubsection{Star 7}
This star has been previously found to be likely unbound and characterized in \cite{Bromley_2018} and \cite{Du_2018} in accord with our result; however, there is disagreement over the origin. \cite{Du_2018}, using LAMOST abundances, finds an [$\alpha$/Fe] of $\sim0.3$ which they use as evidence for this star having an accreted origin. We find $\mathrm{[\alpha/Fe]}\sim 0.4$, typical of the stellar halo.  Caution should be exercised when comparing these measurements in detail. The line selection, or region selection used in \cite{Du_2018} differs from ours as we do not include Ti in our measurements. \cite{Bromley_2018} uses orbit integration to conclude this star may come from the LMC. While we are able to replicate the timing they give for the disc crossing, we do not find a similar result for the LMC close approach. We find no clear dynamical progenitor for this star and conclude it was likely born in the `in situ' stellar halo.


\section{Summary}
\label{summary}
Hyper velocity stars are rare and useful objects to study both due to their unclear origin and potential applications for the understanding of the Galaxy. While first discovered by \cite{brown_2005}, the subfield has seen rapid growth fueled by renewed interest and large scale surveys, in particular the Gaia mission. \cite{Brown_2015} estimated a total of 20 confirmed HVS stars, and \cite{Boubert_2018} find $\sim500$ candidate HVSs. \cite{Boubert_2018} found only 1 of the likely unbound HVS candidates was a late type star.

In this context, we aim to (1) confirm (or not) the HVS status of 16 candidate HVSs taken from the literature, (2) derive their chemical abundance pattern, (3) derive their dynamical properties, and use these pieces of information to (4) constrain their origins. We perform follow-up observations of 16 candidate hyper velocity stars based on the literature to confirm their RVs and measure their chemical abundances. We used a combination of the Tull Echelle Spectrograph on the 2.7m HJST Telescope at the McDonald observatory and the ARCES spectrograph on the 3.5m APO telescope. We find good agreement between the RV measurements from Gaia and our ground based observations.

We use the full 6D kinematic information to assess whether these extreme velocity stars are likely unbound or not on the basis of the Milky Way escape velocity model from \cite{Williams_2017}. We confirm one star (Gaia DR3 source~id 1383279090527227264 ) is very likely unbound, and find 2 (Gaia DR3 source~id 1396963577886583296; Gaia DR3 source~id 1478837543019912064) which might be marginally bound (with the details depending on the exact model of the local escape speed used). The remainder appear high velocity (with $v_{\mathrm{total}} > 300 \ \mathrm{km \ s^{-1}}$, and all but one with $v_{\mathrm{total}} > 350 \ \mathrm{km \ s^{-1}}$) but bound. We use orbit integration to search for a possible dynamic origin of these stars. Between the orbital trajectories of the (marginally) unbound HVS and known globular clusters \citep{galpy_mw_gc} and satellite galaxies \citep{galpy_mw_sat}, we attempt to determine the progenitor. We find that none of the marginally bound or unbound sources have a clear progenitor.


We measure chemical abundances for up to 22 species. These elements are Mg, Si, Ca, Ti ($\alpha$ \ group), Fe, Cr, Ni, Co, Sr, Mn (Fe-Peak), Na, Al, V, Cu, Sc (odd-Z group), and Sr, Y, Zr, Ba, La, Nd, Eu (Neutron capture). These elements span the main nucleosynthetic families. We find our sample is largely consistent with the abundance trends for the inner halo (see e.g., Figures \ref{fig:combined_xfe}). They do not appear to originate from globular clusters, the LMC, or the Galactic center. Unlike \cite{Henrique_2022} and \cite{quispe_2022} we do not find any stars chemically consistent with the low [$\alpha$/Fe] typical of accreted systems. There are three possible causes for this difference: 1) Small number statistics/chance, 2) Differences in abundance analysis, 3) Differences in target selection of HiVel star candidates. For 1) it is possible with small (N$\sim$10--20 stars), that we, by chance sample, different populations of HiVel stars. Additionally, the stellar parameter and abundance analysis methods are different between various literature which could lead to an differences. Finally,  the target selection from this study uses a combination of HVS candidates from four sources as described in Section \ref{target_selection}. \cite{Henrique_2022} also uses \cite{Hattori_2018a} in their initial selection. However, \cite{quispe_2022} uses their own selection process and \cite{Henrique_2022} uses \cite{Herzog-Arbeitman_2018} in addition to \cite{Hattori_2018a}. 

The lack of accreted stars in our sample is intriguing in light of recent results such as \cite{Mackereth_2020}, which propose the majority (70\%) of the halo is accreted. It is possible the in situ halo stars we see are formed in the Milky Way and heated from an early accretion event \citep[e.g.,][]{Belokurov_2020}.  A possible origin for these fast moving stars is they are the metal weak tail of the splash distribution. Many of these stars are on retrograde orbits which can be seen in Figure \ref{fig:toomre}. This agrees with the observation from \cite{Belokurov_2020}. [Al/Fe] has been argued to distinguish between accreted and in situ stars \citep[e.g.,][]{hawkins_2015_components, carrillo_2022}. On the basis of our high [Al/Fe] measurements, it is plausible these stars again formed in the Milky Way, however these observations should be taken with caution. Beyond the difficulties with measuring Al in our moderate SNR sample and the differences in Al measurements between studies we compare our data with other studies (see Section \ref{odd_z}), there are also potential problems with comparing IR measurements of Al with optical ones for metal poor stars \citep[e.g.][]{carrillo_2022}.

To our knowledge, Gaia DR3 source~id 1383279090527227264 is one of the first late-type HVS with high resolution spectra and detailed chemical abundances. Our measurements suggest it is a halo star, ruling out several other proposed origin scenarios. Lacking a known progenitor star cluster, we conclude it was likely accelerated from the Galactic Halo. Gaia DR3 is likely to reveal many new late type HVS candidates for follow-up work. 

\section*{Data Availability}
Measured stellar atmospheric parameters and abundances are available online through the Strasbourg astronomical Data Center (CDS). Literature data was accessed through public archives or online data tables. The public archives used were Pan-STARRS DR1~(\url{https://catalogs.mast.stsci.edu/panstarrs/}), Gaia DR2 and DR3~(\url{https://gea.esac.esa.int/archive/}), SDSS-IV~(\url{https://skyserver.sdss.org/dr17/}), SkyMapper DR2~(\url{https://skymapper.anu.edu.au/data-release/dr2/}), AllWISE~(\url{https://wise2.ipac.caltech.edu/docs/release/allwise/}), and 2MASS~(\url{https://www.ipac.caltech.edu/2mass/}). Data for distances were accessed through the CDS (\url{https://cdsarc.cds.unistra.fr/viz-bin/cat/I/352}). Comparison data samples used in the abundance sections were accessed via the CDS.

\section*{Acknowledgements}
TN \& KH  acknowledge support from the National Science Foundation grant AST-1907417 and AST-2108736 and from the Wootton Center for Astrophysical Plasma Properties funded under the United States Department of Energy collaborative agreement DE-NA0003843. This work was performed in part at the Aspen Center for Physics, which is supported by National Science Foundation grant PHY-1607611. This work was also performed in part at the Simons Foundation Flatiron Institute's Center for Computational Astrophysics during KH's tenure as an IDEA Fellow. This project was developed, in part, at the 2022 NYC Gaia Fete, hosted by the Center for Computational Astrophysics of the Flatiron Institute in New York City.
Grant information:
\\
Facilities used: Apache Point Observatory, McDonald Observatory, Simbad
\\
Software used: {\tt astropy} \citep{astropy:2013, astropy:2018},
{\tt NumPy} \citep{2020NumPy-Array}, {\tt iPython} \citep{ipython}, {\tt Matplotlib} \citep{matplotlib}, {\tt Galpy} \citep{Galpy}, {\tt SciPy} \citep{2020SciPy-NMeth}, {\tt Photutils} \citep{phot_utils}, {\tt BACCHUS} \citep{bacchus2016}, {\tt topcat} \citep{topcat}, {\tt iSpec} \citep{ispec}

\bibliography{stars}

\appendix
\section{ADQL Query For Star 15 and Star 16}
\label{appendix:adql_query}
The query used to find two candidate HVS stars targeted for follow-up observations in this work. The cuts are based on high radial velocity in the galactocentric rest frame.

\begin{lstlisting}[deletekeywords={DEC, DECIMAL, DECLARE}]
SELECT *
FROM gaiadr3.gaia_source
WHERE parallax_error/parallax <= 0.1
AND abs(radial_velocity + 8.5*cos(RADIANS(l))*cos(RADIANS(b)) + 
233.38*sin(RADIANS(l))*cos(RADIANS(b)) + 6.49*sin(RADIANS(b))) >= 400
AND pmra_error < 1
AND pmdec_error < 1
AND radial_velocity_error < 5
AND parallax > 0
\end{lstlisting}

\bsp
\label{lastpage}
\end{document}